\documentclass[reprint,
longbibliography,
 amsmath,amssymb,
 aps,
prstab
]{revtex4-1}

\usepackage{graphicx}% Include Figure files
\usepackage{subcaption}
\usepackage{dcolumn}% Align table columns on decimal point
\usepackage{bm}% bold math
\usepackage{xcolor}
\usepackage{natbib}
\usepackage[%
  colorlinks=true,
  urlcolor=blue,
  linkcolor=blue,
  citecolor=blue
]{hyperref}

\usepackage{amsmath}

\usepackage{textcomp} %Textmu

\makeatletter
\newcommand\thefontsize[1]{{#1 The current font size is: \f@size pt\par}}
\newcommand\thefontsizeHere{{The current font size is: \f@size pt\par}}
\makeatother

\begin{document}

\preprint{APS/123-QED}

\title{Implications of Beam Filling Patterns on the Design of Recirculating Energy Recovery Linacs}% Force line breaks with \\
%\title{An investigation of RF instabilities during beam loading of recirculating Energy Recovery Linac}% Force line breaks with \\

\author{S.~Setiniyaz}
\email{s.saitiniyazi@lancaster.ac.uk}
\author{R.~Apsimon}
\email{r.apsimon@lancaster.ac.uk}
\affiliation{Engineering Department, Lancaster University, Lancaster, LA1 4YW, UK }
\affiliation{Cockcroft Institute, Daresbury Laboratory, Warrington, WA4 4AD, UK}

\author{P.~H.~Williams}
\affiliation{Cockcroft Institute, Daresbury Laboratory, Warrington, WA4 4AD, UK}

\date{\today}% It is always \today, today,

\begin{abstract}

Recirculating energy recovery linacs are a promising technology for delivering high power particle beams ($\sim$GW) while only requiring low power ($\sim$kW) RF sources. This is achieved by decelerating the used bunches and using the energy they deposit in the accelerating structures to accelerate new bunches. We present studies of the impact of the bunch packet filling pattern on the performance of the accelerating RF system. We perform RF beam loading simulations under various noise levels and beam loading phases with different injection schemes. We also present a mathematical description of the RF system during the beam loading, which can identify optimal beam filling patterns under different conditions. 
The results of these studies have major implications for design constraints for future energy recovery linacs, by providing a quantitative metric for different machine designs and topologies.

\end{abstract}

\maketitle

%\begin{itemize}
%	\item short intro into recirculating ERLs
%	\subitem citing design studies around the world
%	\item filling patterns and BL patterns
%	\subitem number of filling patterns for an n-turn ERL
%	\subitem encoding of filling patterns and beam loading patterns
%	\subitem explain how some filling patterns are better or worse than others
%	\item beam loading simulation
%	\subitem explain dynamic and static set-points and show which one is better by simulation
%	\subitem show simulation results for different $S/N$ ratio
%	\subitem brute force method to identify optimal patterns
%	\subitem explain that the different set-points have different Figure of merit for optimal patterns 
%	\item optimal fillingand BL patterns
%	\subitem explain that brute force isn't improved with extra computing power
%	\subitem study properties and characteristics of optimal (or potentially optimal) patterns
%	\subitem describe method for finding optimal patterns
%	\item results
%	\subitem use simulations to complement and verify studies
%	\subitem RF stability for odd and even pass ERLs
%	\subitem key implications of these results
%	\item acknowledgments
%\end{itemize}

\section{Introduction}
\subsection{Introduction into ERLs}
There is an increasing interest in Energy Recovery Linacs worldwide due to their unique promise of combining the high-brightness electron beams available from conventional linacs with the high average powers available from storage rings. Applications requiring this step-change in capability are coming to the fore in a wide variety of fields, for example high energy particle physics colliders~\cite{AbelleiraFernandez2012}, high luminosity colliders for nuclear physics~\cite{Accardi2016}, free-electron laser drivers for academic and industrial purposes~\cite{Socol2013,Socol2011}, and inverse Compton scattering sources~\cite{Shimada2010,Hayakawa2010}. The first high average power application demonstrated on an ERL was the multi-kW lasing of the JLab IR-FEL~\cite{Neil_PRL_ERL_2000}.

\par Historically, an effective method to cost-optimise an electron linac (where beam dynamics restrictions allow) is to implement recirculation~\cite{York1987,Williams2011}, i.e. accelerating the beam more than once within the same RF structures. Analogously, one may implement recirculation in an ERL, accelerating {\it and} decelerating within the same structures. This has been successfully demonstrated in the normal-conducting Novosibirsk infrared FEL~\cite{Shevchenko2016}. There are a number of GeV scale user facilities proposed that are therefore based upon recirculating superconducting ERLs~\cite{AbelleiraFernandez2012,Angal-Kalinin2018,Ptitsyn:2016ckp}, and two test facilities are currently attempting such a multi-turn ERL demonstration~\cite{Hoffstaetter:2017ipp,Arnold:2018ilm}. 

\par It is thus timely to explore the implications of this relatively new accelerator class. Unlike a linac or storage ring, there is large number of degrees of freedom in the basic accelerator topology. For example one may choose a dogbone or racetrack layout, subsequent accelerating pass may be transported in common or separate beam transport, and decelerating passes may be transported pairwise with their equivalent accelerating beam in common or separate transport~\cite{Williams:FLS2018-THA1WA04,Douglas:IPAC2018-THPMK106,WilliamsLHEC2018}. 

\par In this article we explore the consequence of these choices on the most important aspect of an ERL-based user facility, the RF stability. Specifically, we consider all possible beam filling patterns in an N-pass recirculating ERL and their interaction with the accelerator low-level RF control system. We show that there are optimal choices, and note which topologies allow these optima to be chosen. 

\par It is vital that this analysis is performed during the design stage of an ERL-based facility as it fixes the pass-to-pass
path length required in the recirculation transport at the scale of multiples of the fundamental RF wavelength, typically many metres, therefore any path length variability built in to allow pass-to-pass RF phase variation cannot correct for this macro scale requirement. Similarly, transverse phase advance manipulations that are capable of mitigating BBU thresholds~\cite{Douglas2006} would not be effective against sub-optimal filling pattern generated instabilities.

\par We first introduce beam filling and beam loading patterns, and describe how they affect cavity voltage. We then describe an analytical model of beam loading and use this to make predictions about the system. The next section describes beam loading simulations while varying different parameters such as the signal-to-noise ratio ($S/N$) and synchronous phase. We will expand these studies to sequence preserving scheme in the section \ref{section:4} and compare all the simulations results in the section \ref{section:5}.  

\subsection{Filling patterns}
\begin{figure}
  \centering
  \includegraphics[width=50mm]{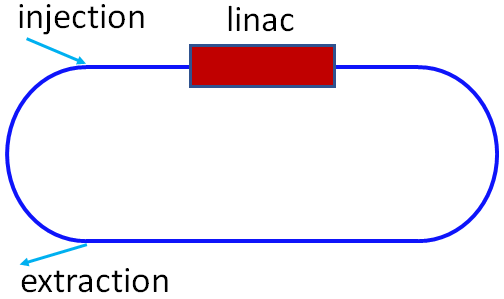}
  \caption{Simple recirculating linac diagram.}
  \label{fig:SimpleERL}
\end{figure}

\begin{figure}
  \centering
  \includegraphics[width=65mm]{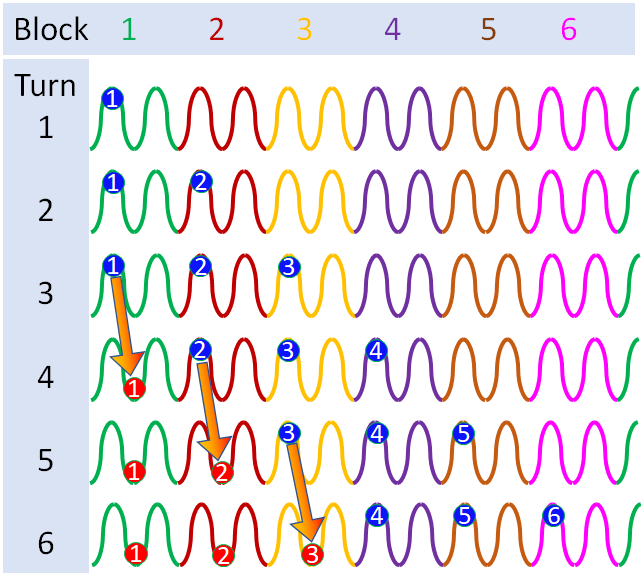}
  \caption{Filling of recirculating linac with filling pattern [1~2~3~4~5~6]. Blue/red bunches are accelerated/decelerated. Phase flips at 3rd turn.}
  \label{fig:filling}
\end{figure}

\begin{figure}
	\centering
	\includegraphics[width=50mm]{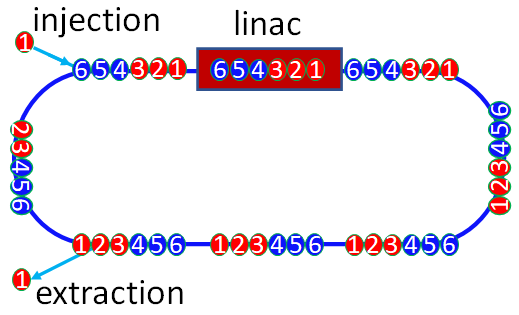}
	\caption{Filling of ERL by multiple bunch packets.}
	\label{fig:packets}
\end{figure}

In this article, we note that the topology of the recirculating ERL can impact the filling pattern or ordering of the bunches. We start with a simple recirculating ERL with single arc on two sides as shown in Fig.~\ref{fig:SimpleERL} and discuss more complex setup later on. 
%The injected bunches in this simple ERL maintain their order until the extraction.
%Throughout this paper, we shall refer to ``injection'' as the process of transporting the bunches from the injection line to the ERL main ring; similarly, ``extraction'' will refer to the process of extracting the bunches from the ring and transporting them to the beam dump. 
%As an example, 
We consider a 6-turn ERL with 3 acceleration and 3 deceleration turns. 
%If all the bunches are injected in one turn, the cavity voltage will decrease or increase drastically, compared to the case where only one bunch is injected per turn. 
In order to minimise cavity voltage fluctuations, we allow for spacing between injected bunches which become filled by bunches on subsequent passes. Here we elucidate the exact choices in which that process occurs. As an example Fig.~\ref{fig:filling} shows 3 decelerating bunches followed by 3 accelerating ones. The accelerating bunches take energy from the cavity, thus decreasing the cavity voltage and vice versa, therefore mixing them can minimize cavity voltage fluctuation. The 6 bunches form what we term a “bunch packet”. Bunch packets are repeated and fill up the ERL as shown by the diagram in Fig.~\ref{fig:packets}.
As we mix bunches executing different turns into bunch packets we emphasise that ``injection'' only refers to the process of transporting a bunch from the injection line to the ERL main ring; similarly, ``extraction'' refers to the process of extracting a bunch from the ring and transporting them to the beam dump. Therefore a set of injected bunches do not pass through the linac as one, they are always mixed with bunches executing turns in the ERL ring.
The ``bunch number" is the order in which bunches are injected into a bunch packet over $N$ turns, for example bunch 1 (or 1st bunch) is injected on turn 1, bunch 2 (or 2nd bunch) is injected as bunch 1 executes turn 2 and so on.
%In the second scenario, the accelerating and decelerating bunches are alternated as shown in Fig.~\ref{fig:filling}. 
%Here, 3 decelerating bunches are followed by 3 accelerating ones. The acceleration takes energy from the cavity and thus decrease cavity voltage and vise versa. So, mixing accelerating bunches with decelerating bunches can minimize cavity voltage fluctuation. These 6 bunches form a bunch packet. Bunch packets are repeated and fill up the ERL as shown by the diagram in Fig.~\ref{fig:packets}. 
During the operation, one bunch per packet per turn is extracted and replaced by a new bunch. Usually, not all the RF cycles are filled by bunches, but one bunch is located at the start of a block of $M$ otherwise unoccupied RF cycles. These $M$ RF cycles we call the ``intra-packet block''. In a $N$-turn ERL, 1 bunch packet thus occupies $M\times N$ RF cycles. In the packet illustrated in Fig~\ref{fig:filling} each intra-packet block is coloured uniquely.

%Usually, not all the RF cycles are filled by bunches, but one bunch occupies $N$ RF cycles. These $N$ RF cycles is called a ``intra-packet block''. In a $M$-turn ERL, 1 bunch packet occupies $M$ RF blocks. 

\begin{figure}
	\centering
	\includegraphics[width=65mm]{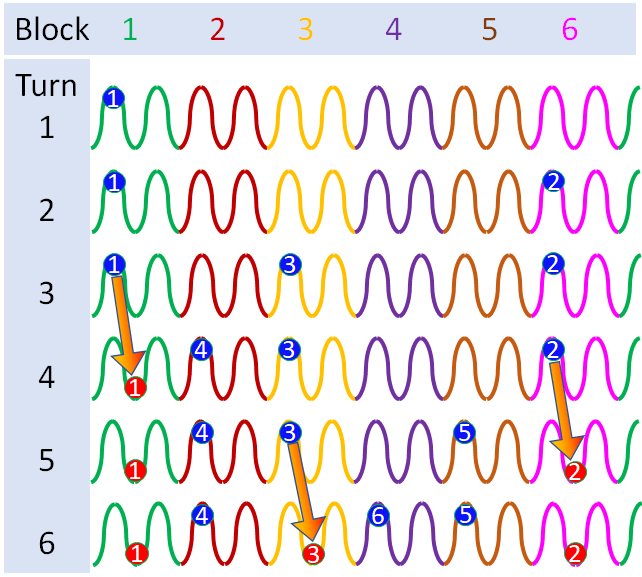}
	\caption{Filling of recirculating linac with filling pattern [1~4~3~6~5~2].}
	\label{fig:filling2}
\end{figure}

%Firstly, It is necessary to define various quantities for use in our analysis. The ``bunch number'' is the order in which bunches are injected into a bunch packet over $N$ turns, for example bunch 1 (or 1st bunch) is injected on turn 1, bunch 2 (or 2nd bunch) is injected on turn 2 and so on. 
We can give notation of filling pattern by describing which bunch goes to which intra-packet block. The number indicates the bunch number and its position in the vector indicates the intra-packet block number. The filling pattern of Fig.~\ref{fig:filling} is a 6-element vector [1~2~3~4~5~6]. Filling pattern [1~4~3~6~5~2], for example, describes filling depicted in Fig.~\ref{fig:filling2}.

Here we attempt a step-by-step explanation of packet construction. 
%We assume flexible injection timing that bunches can be injected with changing time intervals. 
We assume a flexible injection timing, such that we can insert a small delay of less than the regular pulse spacing (but still a multiple of fundamental RF) with a regular superperiod. Such capability would be novel, though not unfeasible, within the photoinjector laser system.
Please refer to Figs.~\ref{fig:filling}, \ref{fig:packets} and \ref{fig:filling2}: We start by injecting all the bunches labeled 1. In Fig.~\ref{fig:packets}, we see that we can fit 8 packets into the ERL (the number of packets in the ring are arbitrarily chosen), so this accounts for the first 8 bunches from the injector. This completes turn 1 in Fig.~\ref{fig:filling} (or Fig.~\ref{fig:filling2}) (which both shows only one of the 8 packets). The ninth bunch from the injector becomes the first “bunch 2” on the second line of Fig.~\ref{fig:filling} or Fig.~\ref{fig:filling2}. In the case of Fig.~\ref{fig:filling} the bunch 2 is injected in to “intra-packet block” number 2. In the case of Fig.~\ref{fig:filling2} the bunch 2 is injected in to “intra-packet block” number 6. 
%This difference between Fig.~\ref{fig:filling} and Fig.~\ref{fig:filling2} is accomplished with a change of the time interval separating the 8th and 9th bunches from the injector.
This difference between Fig.~\ref{fig:filling} and Fig.~\ref{fig:filling2} is accomplished using the aforementioned flexible timing feature of the photoinjector laser by extending the time interval separating the 8th and 9th bunches..
The next 7 injected bunches fill up the other “bunch 2” spaces in the other packets. Following through, the 17th bunch from the injector thus becomes “bunch 3” in the packet, in both Figs.~\ref{fig:filling} and \ref{fig:filling2} this is placed in block number 3. In this way we build up either [1 2 3 4 5 6] for Fig.~\ref{fig:filling}, or [1 4 3 6 5 2] for Fig.~\ref{fig:filling2}. We call patterns constructed in this method "First-In-First-Out" (FIFO) patterns as the order of the bunches in the packet remains constant. 

Another way to construct filling patterns is recombination using different path lengths with a fixed injection time interval. In this case, the turn number of the bunches in the packet does not change. Therefore, we name it Sequence Preserving (SP) scheme. We will discuss it in more details in later sections. A point we wish to emphasize for SP scheme is that because choosing between these two filling patterns implies differences in the path lengths of many RF cycles for each individual turn, this choice is a design parameter during machine construction.

We will also use ``pattern number'' for brevity to indicate 120 filling patterns of 6-turn ERL. The pattern number $i$ is used to indicate 120 permutations of [2~3~4~5~6] and related to the filling pattern $F_i$ as

\begin{equation}
\begin{split}
    &F_{1} = [1~2~3~4~5~6], \\
    &F_{2} = [1~2~3~4~6~5], \\
    &... \\
    &F_{120} = [1~6~5~4~3~2].
\end{split}
\label{eq:ustored0}
\end{equation}

%Similarly, filling pattern [1~4~3~6~5~2] will describe filling shown in Fig.~\ref{fig:filling2}. 
\noindent As there are many bunch packets in a ring, without losing the generality we can name intra-packet block of the 1st bunch as the 1st block, i.e. the 1st bunch will always be in the 1st intra-packet block. %When referring to an $N$-pass machine, ``pass'' indicates the number of accelerating passes in the ERL, which may be any number, the total number of ``turns'' will therefore be $2N$.

\subsection{Cavity voltage calculation}
As the bunches pass through the linacs, they are either accelerated or decelerated by the RF field in the cavity. In doing so, energy is either put into or taken out of the cavity. The cavity voltage $V_{cav}$ is related to the stored energy $U_{stored}$ as

\begin{equation}
    U_{stored} = \frac{V_{cav}^{2}}{\omega\left(\frac{R}{Q}\right)}
    \label{eq:ustored},
\end{equation}

\noindent with $\frac{R}{Q}$ being shunt impedance of the cavity divided by its Q-factor. For an accelerating cavity, the change in stored energy from a particle bunch passing through is

\begin{equation}
    \delta{U}_{stored}=\frac{2V_{cav}\delta{V}_{cav}}{\omega\left(\frac{R}{Q}\right)}=-q_{bunch}V_{cav}.
    \label{eq:ustored1}
\end{equation}

\noindent Therefore, the change in cavity voltage from beam loading is given as

\begin{equation}
    \delta{V}_{cav}=-\frac{q_{bunch}}{2}\omega\left(\frac{R}{Q}\right)\cos{\left(\phi\right)},
    \label{eq:ustored2}
\end{equation}

\noindent where $\phi$ is the phase difference between the bunch and the RF and $q_{bunch}$ is the bunch charge. In general, the bunches will not necessarily pass through the cavity on-crest (maximum field) or on-trough (minimum field). When dealing with RF fields, it is convenient to consider the field as a complex number, where only the real part can interact with the beam at any moment in time. Indeed this implies that beam loading can only change the real component of the cavity voltage for any given phase.

\begin{figure}
  \centering
  \includegraphics[width=85mm]{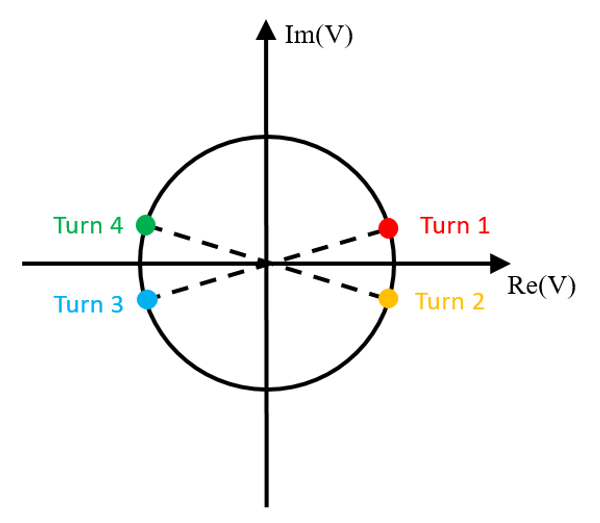}
  \caption{A diagram to show the complex voltages of four bunches in a 4-turn ERL.}
  \label{fig:BL_ERL}
\end{figure}

In order for a recirculating ERL to operate stably over time, we require that the vector sum of the cavity voltage experienced by each bunch in a bunch packet must equal zero, as shown Fig.~\ref{fig:BL_ERL}. If this is not the case, then there will be a net change in stored energy in the cavity each bunch packet, reducing the overall efficiency of the ERL.

For now, we will neglect the phase of the bunches and only consider voltages as real numbers for brevity in the following mathematical description. Later we will consider off-crest beam loading cases by replacing binary notation with complex notation, i.e. by replace ``1'' and ``0'' by $e^{i\phi}$ and $e^{-i\phi}$. We define a recirculating ERL to be at `steady state' when all intra-packet blocks in the machine are occupied. In this case, on any given turn, half the bunches in the packet pass through the cavity at accelerating phases and half at decelerating phases. As cavity voltage experienced by all bunches in the packet sum to zero, there is no net energy gain or loss over bunch packet.

If we neglect the phase of the bunches and only consider bunches passing through the cavity on-crest and on-trough, then the change in cavity voltage due to beam loading from a bunch is simply $\pm\frac{q_{bunch}}{2}\omega\left(\frac{R}{Q}\right)\cos{\left(\phi\right)}$, from Eq.~\ref{eq:ustored2}. Therefore, in this case, every time a bunch passes through a linac, the cavity voltage is incremented or decremented by a fixed amount.

\subsection{Beam loading pattern}
\begin{figure}
	{\scalebox{0.5} [0.5]{\includegraphics{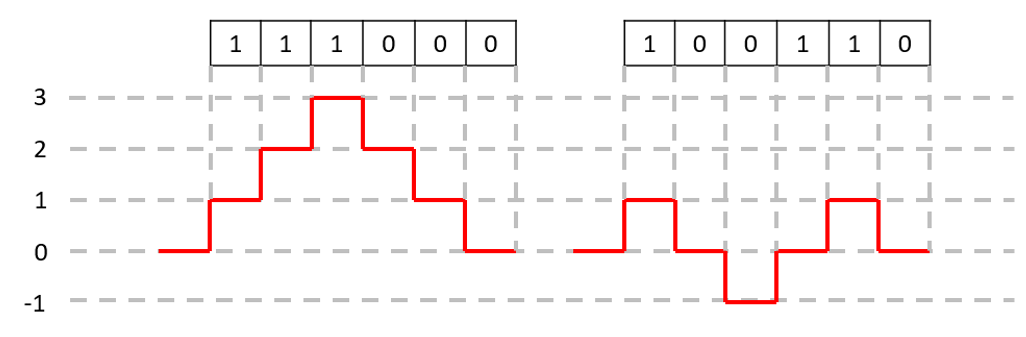}}}\\
	\caption{A diagram to show how the beam loading pattern translates into a change in cavity voltage over time.}
	\label{fig:BL_patterns}
\end{figure}

Let us consider a 6-turn ERL. Table~\ref{tab:bl_patt} shows how the beam loading pattern changes turn-by-turn for the filling patterns [1~2~3~4~5~6], [1~4~3~6~5~2], and [1~4~5~2~3~6]. If we use ``0'' and ``1'' to denote accelerated and decelerated bunches, respectively, we get beam loading patterns as shown in Table~\ref{tab:bl_patt}. The accelerating bunches reduce the voltage in the cavity and vise versa. Now that we have defined the bunch filling pattern and showed how this is associated with a unique sequence of beam loading patterns, we should understand how this beam loading pattern affects the cavity voltage. Fig.~\ref{fig:BL_patterns} shows how the beam loading pattern can be translated into a change in cavity voltage. 

%\begin{figure}
%	\centering
%	\includegraphics[width=85mm]{ColorTable.png}
%	\caption{Filling patterns and associated beam loading patterns.}
%	\label{tab:bl_patt}
%\end{figure}

\begin{table}
	\caption{Filling patterns and associated beam loading patterns.}
	\begin{ruledtabular}
		\begin{tabular}{llll}
%			filling pattern No. & 1 & 60 & 61 \\	
			filling pattern & 1 2 3 4 5 6 & 1 4 3 6 5 2 & 1 4 5 2 3 6 \\	
			\hline
            turn 1 & 0 & 0 & 0 \\
            turn 2 & 0~0  & 0~\textcolor{white}{1~1~0~1}~0  & 0~\textcolor{white}{1~1}~0\textcolor{white}{~1~0} \\
            turn 3 & 0~0~0  & 0~\textcolor{white}{1}~0~\textcolor{white}{1~1}~0  & 0~\textcolor{white}{1~1}~0~0\textcolor{white}{~1} \\
            turn 4 & 1~0~0~0  & 1~0~0~\textcolor{white}{1~1}~0  & 1~0~\textcolor{white}{1}~0~0\textcolor{white}{~1} \\
            turn 5 & 1~1~0~0~0  & 1~0~0~\textcolor{white}{1}~0~1 & 1~0~0~1~0\textcolor{white}{~1} \\
            turn 6 & 1~1~1~0~0~0  & 1~0~1~0~0~1 & 1~0~0~1~1~0 \\
            turn 7 & 0~1~1~1~0~0  & {0~1}~{1~0~0~1}  & {0~1}~{0~1~1~0} \\
            turn 8 & 0~0~1~1~1~0  & {0~1~1~0}~{1~0}  & {0~1}~{1~0}~{1~0} \\
            turn 9 & 0~0~0~1~1~1  & {0~1}~{0~1}~{1~0}& {0~1~1~0}~{0~1} \\
            turn 10 & 1~0~0~0~1~1 & {1~0}~{0~1~1~0}  & {1~0}~{1~0~0~1} \\
            turn 11 & 1~1~0~0~0~1 & {1~0~0~1}~{0~1}  & {1~0}~{0~1}~{0~1} \\
            turn 12 & 1~1~1~0~0~0  & 1~0~0~1~1~0  & 1~0~1~0~0~1 \\
        \end{tabular}
     \end{ruledtabular}  
     \label{tab:bl_patt}
\end{table} 

For an ERL at steady state, the definition of ``block 1'' is arbitrary and can be one of $N$ choices in a $N$-turn ERL; therefore, there are $\left(N - 1\right)!$ unique bunch filling patterns for a $N$-turn ERL. A 6-turn ERL can have 120 unique filling patterns. Each of these filling patterns is associated with a unique sequence of beam loading patterns. Beam loading patterns changes turn by turn and are periodic over $N$ turns, as shown in Table~\ref{tab:bl_patt}. 

\begin{figure}
		{\scalebox{0.35} [0.35]{\includegraphics{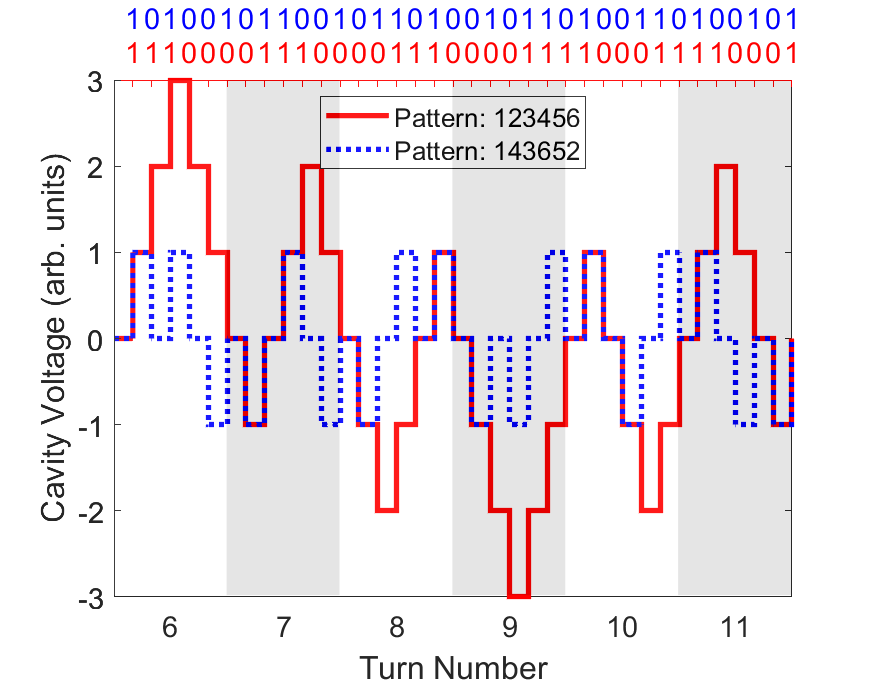}}}\\
%       {\scalebox{0.35} [0.35]{\includegraphics{6turnBLPatt60N1_2.png}}}\\
	\caption{Comparison of cavity voltage change by two different filling patterns over 6-turns.}
	\label{fig:6Turn_BL_patterns}
\end{figure}

Fig.~\ref{fig:6Turn_BL_patterns} shows beam loading patterns of two filling patterns over 6-turns. The red beam loading pattern has larger cavity voltage fluctuation than blue one. This shows some filling patterns cause larger disturbances to the cavity voltage and RF system of the ERL than others. For a 6-turn ERL, we can evaluate the RF jitters associated with a specific beam filling pattern and use this to identify which patterns are optimal. In Table~\ref{tab:bl_patt}, the beam loading increments have been normalised to $\pm1$ rather than $\pm\frac{q_{bunch}}{2}\omega\left(\frac{R}{Q}\right)\cos{\left(\phi\right)}$ for brevity and clarity. For the remainder of the article, we will continue to use a normalised beam loading to help the reader understand the methodology.

Once a list of all unique filling patterns is defined, we can determine the associated sequence of beam loading patterns, using the method described in Table~\ref{tab:bl_patt}. To determine the normalised change in cavity voltage, we simply calculate the cumulative sum of the beam loading sequence. We define a specific filling pattern as $F_{i}$, the associated beam loading pattern as $B\left(F_{i}\right)$ and the normalised change in cavity voltage as $\delta{V}$ given as

\begin{equation}
    \delta{V} = \text{cusum}\left(B\left(F_{i}\right)\right)=\sum_{j=1}^{k}{B_{j}\left(F_{i}\right)}.
    \label{eq:cusum}
\end{equation}

\noindent We can use $\delta{V}$ to estimate the RF stability performance of all patterns.

\subsection{Low level RF system}
\begin{figure}
		{\scalebox{0.3} [0.3]{\includegraphics{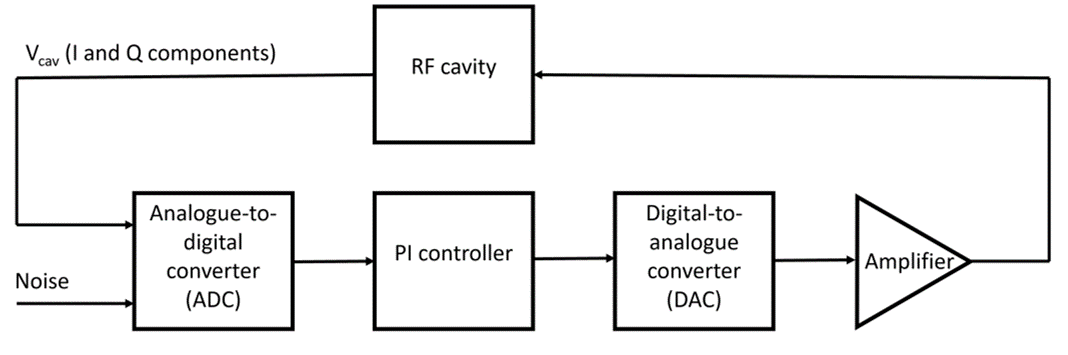}}}\\
%       {\scalebox{0.35} [0.35]{\includegraphics{6turnBLPatt60N1_2.png}}}\\
	\caption{A block diagram of the modelled LLRF system and the feedback loop.}
	\label{fig:LLRFBlock}
\end{figure}

For the Low level RF (LLRF) system, we model the system as shown in Figure~\ref{fig:LLRFBlock}. The cavity voltage (given as I and Q components) is added to a Gaussian distributed noise (also I and Q), whose standard deviation is defined by the $S/N$; we treat this as the only source of noise in the system, rather than including realistic noise at each component of the LLRF controller. This is then passed through a 16-bit analogue-to-digital converter (ADC), before a PI-control algorithm is implemented to regulate amplitude and phase. The PI correction algorithm also applies limits to the range of values to model the power limits on the amplifier. The amplifier and digital-to-analogue converter (DAC) is modeled as a resonant circuit with a bandwidth defined by the closed-loop bandwidth.

We model LLRF system as a proportional-integral (PI) controller~\cite{DEXTER201762,Sjobak:2016ruf,Baudrenghien}. In the PI controller, the LLRF system first calculates the error $u$ voltage, which is difference between actual cavity voltage $V_{measured}$ with set-point voltage $V_{set}$ 

\begin{equation}
u = V_{measured} - V_{set}.
    \label{eq:dV}
\end{equation}

\noindent Then, two types of corrections are made, namely the proportional $V_{pro}$ and integral term corrections $V_{int}$. The proportional term correction is calculated based on the previously measured $dV$ and proportional gain $G_p$, given as 
\begin{equation}
V_{pro} = G_{p} u.
    \label{eq:P-term}
\end{equation}

\noindent The integral term correction is calculated integrating over on all the previously measured $dV$ and integral term gain $G_i$, given as

\begin{equation}
V_{int} = G_i \int_{0}^{t} udt = G_{i}\sum_{n}{u_{n}\delta{t}},
    \label{eq:P-term}
\end{equation}

\noindent where $t$ is the time measurement took place. The proportional and integral term corrections address fast and slow changes, respectively. 
%The $V_{int}$ is ignored in Eq.~\ref{eq:PampAve2}, because integral of $V_b$ and $V_n$ over long time is zero. Here, $G_p$ is the proportional gain of LLRF system. The noise in the cavity voltage is amplified by LLRF system by $G_{p}$ times. 
The set-point voltage can be constant (static set-point) or can change over time (dynamic set-point). A dynamic set-point can be useful in order to improve RF stability in a recirculating ERL because it prevents the LLRF system from competing with the beam loading voltage in the cavity. If the LLRF feedback system can adjust its set-point voltage according to the anticipated beam loading, then it has a ``dynamic set-point'' voltage. In this case, the feedback system only amplifies noise. If the set-point is static, LLRF system will treat beam loading as noise and amplify it as well.

% In this case, $V_b$ is multiplied to $G_{p}$ as well. % and $G_p$ is multiplied to $V_{n}$ only, as in Eq.~\ref{eq:PampAve2}.
\section{Analytical model}
\subsection{Variations in cavity voltage}
If we consider the effects of beam loading and noise, the cavity voltage, $V_{cav}$, can be expressed as:

\begin{equation}
    V_{cav} = V_{0} + V_{b} + V_{n},
    \label{eq:Amod1}
\end{equation}

\noindent where $V_{0}$ is the steady state cavity voltage, which we will assume to be time-independent, $V_{b}$ is the voltage contribution due to beam loading, and $V_{n}$ is the voltage contribution due to all noise sources in the system. We shall assume that noise originates from the electronics in the low-level RF system (LLRF), which in turn introduces noise to the cavity voltage. How the noise propagates through the RF system depends on the behaviour of the LLRF system as well as the beam loading patterns, but the noise voltage in the cavity can be defined as

\begin{equation}
    \sigma_{V_{n}} = \frac{\alpha_{RF}\lvert V_{0} \rvert }{S/N},
    \label{eq:Amod2}
\end{equation}
\noindent where $S/N$ is the voltage signal to noise ratio and $\alpha_{RF}$ is a constant of proportionality, which depends on the parameters of the system. From Eq.~\ref{eq:Amod1}, we can obtain an expression for the cavity voltage squared:

\begin{equation}
    V_{cav}^{2} = V_{0}^{2}+V_{b}^{2}+V_{n}^{2}+2V_{0}V_{b}+2V_{0}V_{n}+2V_{b}V_{n}.
    \label{eq:Amod3}
\end{equation}

\noindent We shall assume that $V_{b}$ and $V_{n}$ are independent variables and that $V_{0}$ is constant, therefore, from Eq.~\ref{eq:Amod1} and \ref{eq:Amod3}, we obtain expressions for the mean and standard deviation of the cavity voltage.

\begin{equation}
    \begin{array}{l}
        \langle{V_{cav}}\rangle = V_{0} + \langle{V_{b}}\rangle + \langle{V_{n}}\rangle \\ \\
        \sigma_{V_{cav}} = \sqrt{\langle V_{cav}^2 \rangle - \langle V_{cav} \rangle^2}
    \end{array}
    \label{eq:Amod4}
\end{equation}

\noindent If $V_{b}$ and $V_{n}$ have zero mean, then Eq.~\ref{eq:Amod4} produces the expected result that $\langle{V_{cav}}\rangle = V_{0}$. Because noise and beamloading is independent, 

\begin{equation}
    \langle V_{b} V_{n} \rangle = \langle V_{b}\rangle \langle V_{n}\rangle.
    \label{eq:Amod5}
\end{equation}

\noindent Therefore, 

\begin{equation}
     \sigma_{V_{cav}} = \sqrt{\sigma_{V_{b}}^{2} + \sigma_{V_{n}}^{2}}.   
    \label{eq:Amod6}
\end{equation}

\noindent From Eqs.~\ref{eq:Amod2} and \ref{eq:Amod6}, we can express the noise on the cavity voltage as

\begin{equation}
     \sigma_{V_{cav}} = \sqrt{\sigma_{V_{b}}^{2} + \alpha_{RF}^{2}\frac{V_{0}^{2}}{\left(S/N\right)^{2}}}.   
    \label{eq:Amod7}
\end{equation}

\begin{figure}
		{\scalebox{0.5} [0.5]{\includegraphics{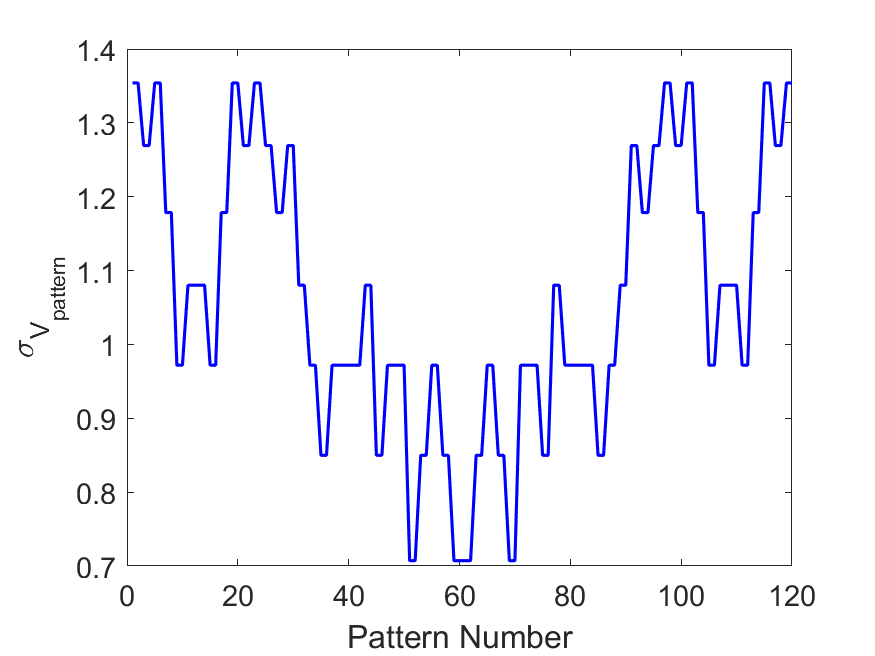}}}\\
	\caption{The RMS fluctuation of the normalized beam loading pattern of 6-turn ERL.}
	\label{fig:Perdicted-SigVb}
\end{figure}

\noindent The $\sigma_{V_{b}}$ is pattern specific, and depends on topology of the ERL as well as the expected beam jitters. The voltage fluctuation due to the beam loading and given by

\begin{equation}
\sigma_{V_{b}} =  \sigma_{V_{pattern}}\delta V
    \label{eq:Amod8}
\end{equation}

\noindent where $\sigma_{V_{pattern}}$ is RMS fluctuation of the normalized beam loading pattern over all turns of the machine. The $\sigma_{V_{pattern}}$ for all 120 patterns is shown in Fig.~\ref{fig:Perdicted-SigVb} for a 6-turn ERL, where we have assumed a FIFO schemes, where the order of the bunch packet does not change turn by turn. One can see that $\sigma_{V_{pattern}}$ varies by approximately a factor of 2 depending on the choice of filling pattern.

\subsection{Variations in amplifier power}
From \cite{RobPRAB}, the cavity voltage can be determined from an envelope equation

\begin{equation}
\frac{\dot{\mathbf{V}}_{\mathrm{cav}}}{\omega_{0}}+\left[\frac{\omega_{0}^{2}+\omega^{2}}{4Q_{L}\omega^{2}}+j\frac{\omega_{0}^{2}-\omega^{2}}{2\omega\omega_{0}}\right]\mathbf{V}_{\mathrm{cav}} = \frac{j\dot{\mathbf{V}}_{amp}+\omega\mathbf{V}_{amp}}{\omega{Q_{e}}}
	\label{eq:Amod9}.
\end{equation}

\noindent Where $\omega_{0}$ is the resonant frequency of the cavity, $\omega$ is the amplifier drive frequency, $Q_{L}$ and $Q_{e}$ are the loaded and external Q-factors respectively and $P_{amp}$ is the forward power from the amplifier expressed as a voltage as

\begin{equation}
\mathbf{V}_{amp} = \sqrt{2\left(\frac{R}{Q}\right)Q_{e}\mathbf{P}_{amp}}
	\label{eq:Amod10}.
\end{equation}

\noindent If we assume that the cavity is driven at the resonant frequency and that the cavity is at steady state, then from Eq.~\ref{eq:Amod9}, we obtain

\begin{equation}
\mathbf{V}_{cav} = \frac{2Q_{L}}{Q_{e}}\mathbf{V}_{amp}
	\label{eq:Amod11},
\end{equation}

\noindent thus

\begin{equation}
P_{amp} =  \frac{Q_{e}}{8\left(\frac{R}{Q}\right)Q_{L}^{2}}\lvert V_{cav}^2\rvert
    \label{eq:Amod12}.
\end{equation}

\noindent From Eqs.~\ref{eq:Amod3} and \ref{eq:Amod12}, we obtain

\begin{equation}
\begin{aligned}
\langle{P_{amp}}\rangle =  \frac{Q_{e}}{8\left(\frac{R}{Q}\right)Q_{L}^{2}}\left[ V_{0}^{2}+\langle{V_{\beta}^{2}}\rangle+\langle{V_{n}^{2}}\rangle\right. \\
\left.+2V_{0}\langle{V_{\beta}}\rangle+2V_{0}\langle{V_{n}}\rangle+2\langle{V_{\beta}}\rangle\langle{V_{n}\rangle}\right]
    \label{eq:Amod13}.
\end{aligned}
\end{equation}

Note that for the beam loading terms, we now use $V_{\beta}$ rather than $V_{b}$. This is because the LLRF feedback algorithm determines the power required to maintain a stable cavity voltage. If we implement a static set point algorithm, then $V_{\beta}=V_{b}$, if a dynamic set point algorithm is used then $V_{\beta}=\delta{V_{b}}$, which is an error residual when subtracting the expected beam loading voltage from the real value. This error residual depends on pattern number, LLRF algorithm, gains and other factors.

We should note that for the amplifier power, the noise has a simpler relationship to the signal to noise ratio than the noise observed on the cavity voltage (Eq.~\ref{eq:Amod2}) because the noise on the amplifier is the measured noise amplified by the proportional gain of the LLRF, so

\begin{equation}
    \sigma_{V_{n}} = \frac{G_{p}}{S/N}V_{0}
    \label{eq:Amod14}.
\end{equation}

\noindent If we assume that $V_{\beta}$ and $V_{n}$ are independent and zero mean, then Eq.~\ref{eq:Amod13} can be simplified as:

\begin{equation}
\langle{P_{amp}}\rangle =  \frac{Q_{e}V_{0}^{2}}{8\left(\frac{R}{Q}\right)Q_{L}^{2}}\left[ \left(1+\frac{G_{p}^{2}}{\left(S/N\right)^{2}}\right)+\frac{\sigma_{V_{\beta}}^{2}}{V_{0}^{2}}\right]
    \label{eq:Amod15}.
\end{equation}

\noindent By a similar method, we can also determine the standard deviation on the amplifier power as

\begin{equation}
\begin{aligned}
\sigma_{P_{amp}} \approx & \frac{Q_{e}V_{0}^{2}}{8\left(\frac{R}{Q}\right)Q_{L}^{2}}\sqrt{\frac{2G_{p}^{4}}{\left(S/N\right)^{4}}+\frac{2G_{p}^{2}}{\left(S/N\right)^{2}}+\Delta} \\
\Delta = & \frac{\langle{V_{\beta}^{4}}\rangle - \langle{V_{\beta}^{2}}\rangle^{2}}{V_{0}^{4}} + \frac{4\langle{V_{\beta}^{2}}\rangle}{V_{0}^{2}}
    \label{eq:Amod16}.
\end{aligned}
\end{equation}

\noindent For low signal to noise ratios, the first terms dominates, whereas for high signal to noise ratios, we encounter a noise floor due to either beam loading (static set-point) or a residual error (dynamic set-point); this noise floor will be pattern dependent. For the first term, note that it is independent of beam loading pattern and therefore, for lower signal to noise ratios, we expect $\sigma_{P_{amp}}$ to be independent of beam loading pattern.

\section{Beam loading simulation}

The cavity voltage fluctuation can be simulated by simulating beam loading and its interaction with RF system~\cite{RobPRAB}. In this work we have extended beam loading type to accelerating and decelerating. In accelerating mode, voltage changes due to the beam loading is subtracted from cavity voltage and vise versa. 

\subsection{Static and dynamic set-points}

\begin{figure}
	\begin{tabular}{c}
		{\scalebox{0.5} [0.5]{\includegraphics{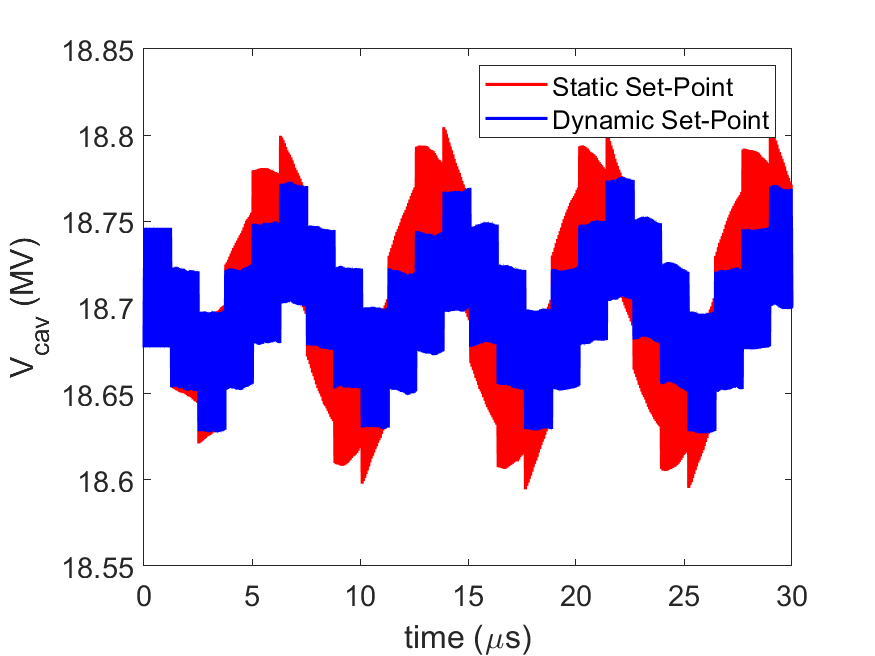}}}\\
		(a) \\
		{\scalebox{0.5} [0.5]{\includegraphics{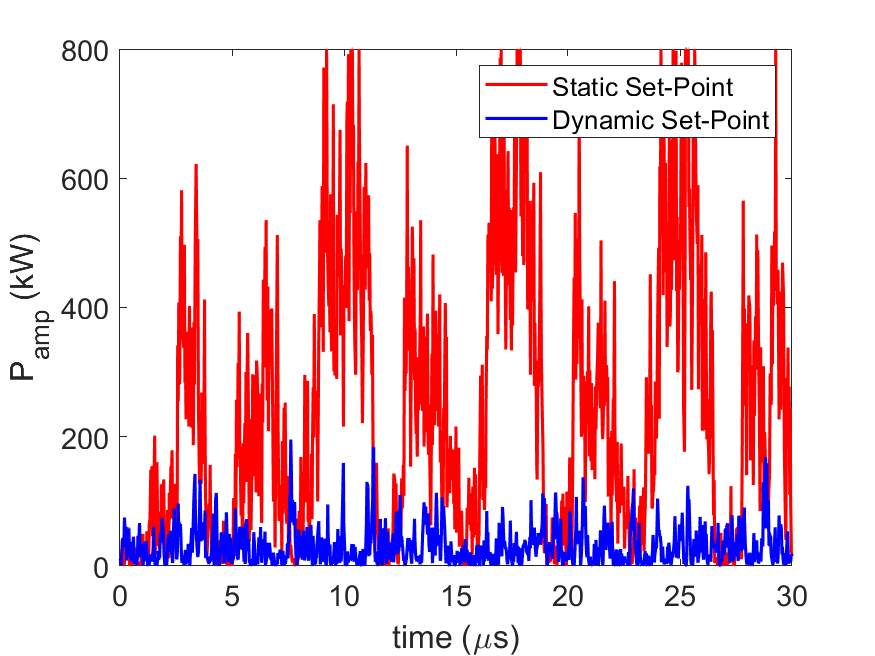}}}\\
		(b) \\
	\end{tabular}
	\caption{Comparison of static and dynamic set-points for filling pattern [1~2~3~4~5~6] when $S/N=7.1\times10^2$. (a) cavity voltage and (b) amplifier power as function of time.}
	\label{fig:stat-dyna-setpoints}
\end{figure}

Before running simulations, it is important to determine the set-point voltage of LLRF system. As we mentioned earlier, there are two types of set-point voltages: dynamic and static set-points. During the beam loading, the cavity voltage fluctuates but the net beam loading of a packet is zero and voltage will return to nominal voltage. So, there is no need for LLRF correction for beam loading. The dynamic set-point is designed to exclude beam loading correction. In static set-point, however, the LLRF system treats beam loading as noise, tries to correct to the oscillatory beam loading, and thus becomes unstable. Therefore, the dynamic set-point is better than static set-point as it creates less cavity voltage fluctuation and requires much less amplifier power. This is also confirmed by simulations shown in Fig.~\ref{fig:stat-dyna-setpoints}. 

\subsection{Simulation parameters}
The simulation parameters are shown in Table~\ref{tab:sim-setup}. We simulated 6-turn ERL, so there are 6 bunches in the packet. The bunch charge was set high to increase the effect of the beam loading and to allow us to explore the behaviour of the RF system under extreme conditions. The circumference is set to 360 m, so number of RF cycles in the ring would be 1200 for a 1~GHz RF frequency. We set 1 intra-packet block is 10 RF cycles, so 20 packets fill up the ring. New bunches replaced old bunches, until total of 96 turns are tracked, which is about 121~$\mu$s time duration. We scanned through all the 120 filling patterns of 6-turn ERL. 

\begin{table}
	\caption{Simulation parameters.}
	\begin{ruledtabular}
		\begin{tabular}{lc}
			\textbf{Machine parameters} & \textbf{value}  \\	
			\hline
            bunch charge $q_{bunch}$ &  18.4 nC \\
            RF cycles per block & 10 \\
            bunches per packet & 6 \\
            number of bunch packets & 20 \\
            circumference & 360 m \\ % = 1200*3e8/freqeuncy 
            revolution time & 1.2 $\mu$s \\ % = circumference/3e8
            number of turns tracked & 96 \\
            tracking time duration & 121 $\mu$s \\  % = 90*1.2 μs 
            \hline
            \textbf{Cavity parameters} &  \\
            \hline
            cavity voltage ($V_{0}$) & 18.7 MV \\
            R/Q & 400  \\
            RF frequency & 1 GHz \\
            \hline
            \textbf{LLRF parameters} &   \\	
			\hline
            latency & 1~$\mu$s \\
            digital sampling rate & 40~MHz \\
            closed-loop bandwidth & 2.5~MHz \\
            proportional controller gain $G_{p}$ & 1000 \\
            integral controller gain $G_{i}$ & 1 \\
            maximum amplifier power & 800~kW \\
        \end{tabular}
     \end{ruledtabular}  
     \label{tab:sim-setup}
\end{table}

\subsection{Simulation results}

\subsubsection{Comparison of optimal and non-optimal patterns}
\begin{figure*}
  \begin{subfigure}[b]{0.4\textwidth}
    \includegraphics[width=\textwidth]{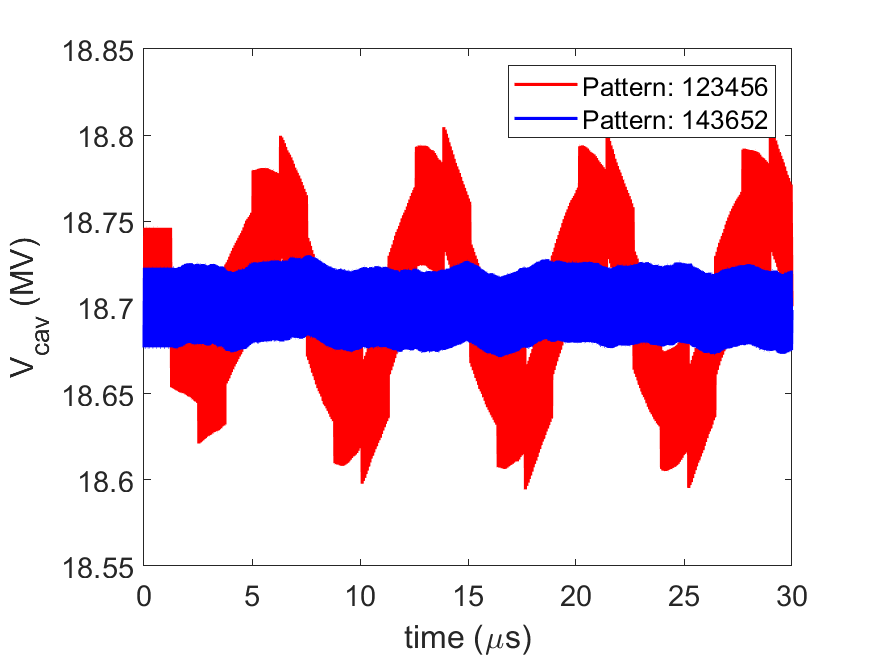}
    \caption{$S/N=7.1\times10^2$}
  \end{subfigure}
  \begin{subfigure}[b]{0.4\textwidth}
    \includegraphics[width=\textwidth]{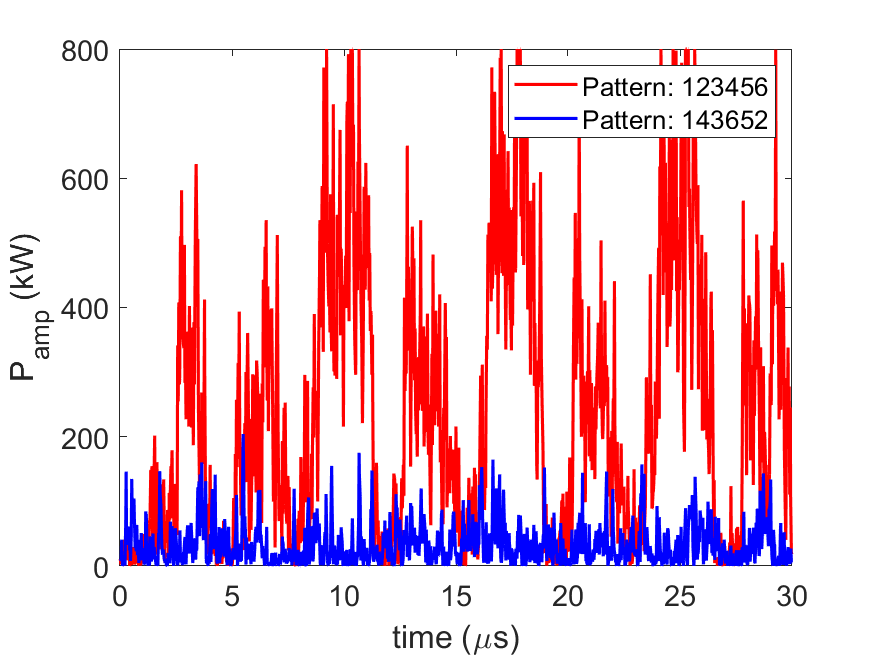}
    \caption{$S/N=7.1\times10^2$}
  \end{subfigure}
    \begin{subfigure}[b]{0.4\textwidth}
    \includegraphics[width=\textwidth]{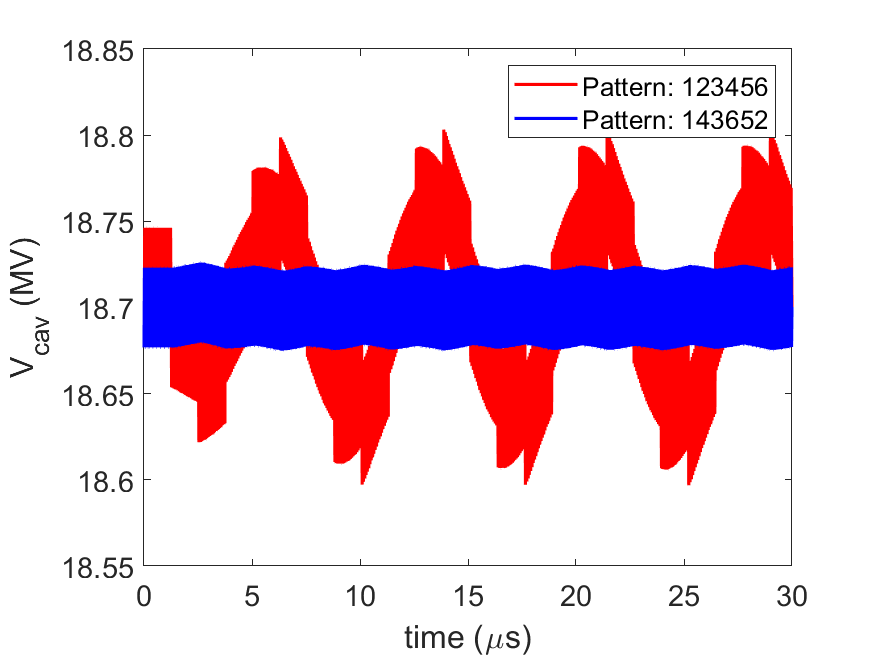}
    \caption{$S/N=7.1\times10^5$}
  \end{subfigure}
  \begin{subfigure}[b]{0.4\textwidth}
    \includegraphics[width=\textwidth]{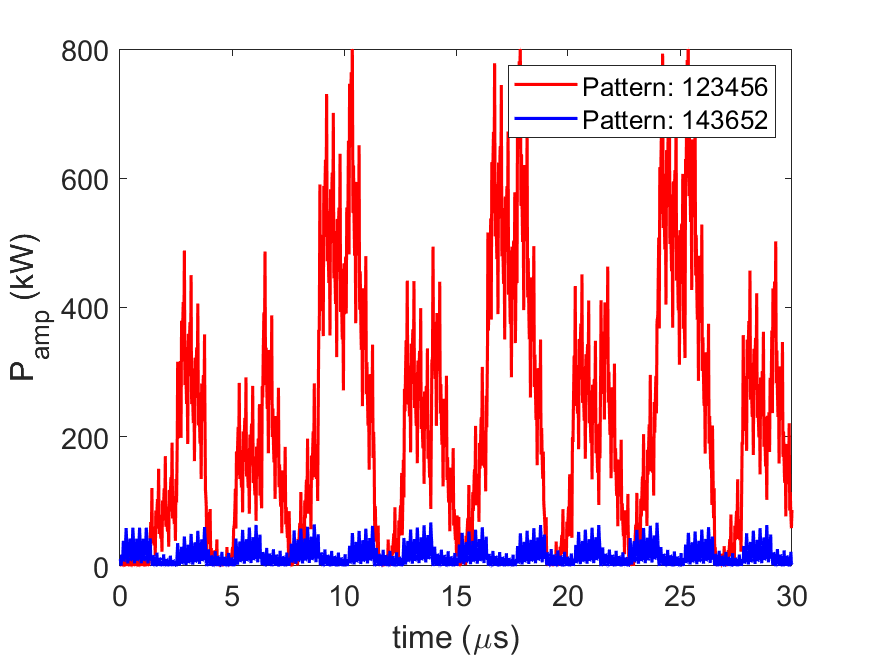}
    \caption{$S/N=7.1\times10^5$}
  \end{subfigure}
  	\caption{Comparison of patterns [1~4~3~6~5~2] and [1~2~3~4~5~6] with dynamic set-point at different $S/N$. (a) and (c) cavity voltage. (b) and (d) amplifier power. (a) and (b) are the results when $S/N=7.1\times10^2$. (c) and (d) are the results when $S/N=7.1\times10^5$.}
  	\label{fig:sim-patt1N60}
\end{figure*}

Firstly, we have looked at the effect of beam loading pattern on the cavity voltage and amplifier power. As show in Fig.~\ref{fig:sim-patt1N60}, the simulation results are shown for an optimal filling pattern [1~4~3~6~5~2] indicated by blue line and a non-optimal pattern [1~2~3~4~5~6] indicated by red line. The optimal pattern is better, because it creates much smaller cavity voltage fluctuations as shown in sub-figures (a) and (c) and requires less amplifier power as shown in sub-figures (b) and (d). The sub-figures (a) and (b) are simulation results when $S/N=7.1\times10^2$ and (c) and (d) are results when $S/N=7.1\times10^5$. Increasing the $S/N$ reduced cavity voltage fluctuation slightly and amplifier power significantly. Simulation results confirmed that certain patterns are better from the perspective of cavity voltage jitters, RF stability, and power requirements. 

\subsubsection{Noise scan}
\begin{figure}
	\begin{tabular}{c}
		{\scalebox{0.5} [0.5]{\includegraphics{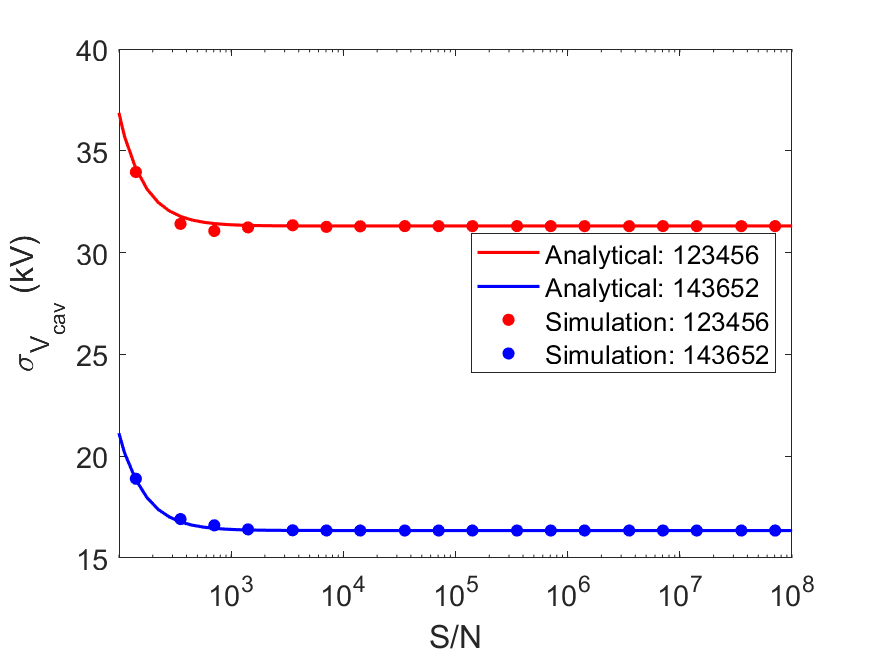}}} \\
		(a)  \\
		{\scalebox{0.5} [0.5]{\includegraphics{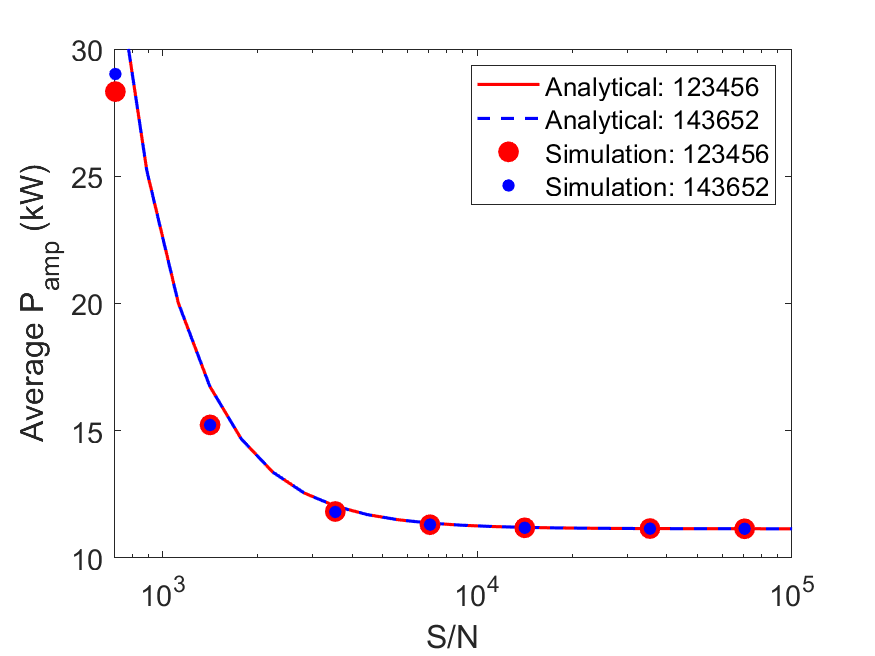}}}\\
		(b)  \\
		{\scalebox{0.5} [0.5]{\includegraphics{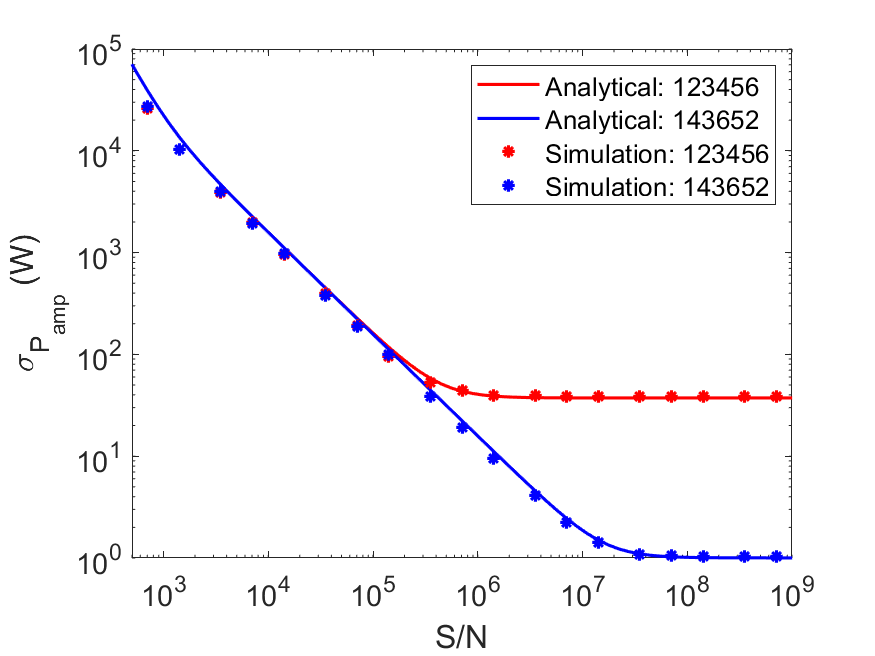}}}\\
		(c)  \\		
	\end{tabular}
	\caption{RMS cavity voltage (a),  average amplifier power (b), and RMS amplifier power (c) as function of $S/N$ for patterns [1~4~3~6~5~2] and [1~2~3~4~5~6].}
	\label{fig:patts-vs-SN}
\end{figure}

We observed the cavity voltage jitters and amplifier power is reduced when $S/N$ is increased. To investigate noise dependence, we have performed simulations with filling patterns [1~4~3~6~5~2] and [1 2 3 4 5 6] by varying $S/N$. The results are shown in Fig.~\ref{fig:patts-vs-SN} for (a) $\sigma_{V_{cav}}$, (b) $\sigma_{P_{amp}}$, and (c) average $P_{amp}$.

In Fig.~\ref{fig:patts-vs-SN} (a), we see that the $\sigma_{V_{cav}}$ is more sensitive to the filling pattern than $S/N$. In other words, $\sigma_{V_{cav}}$ is dominated by filling pattern. $\sigma_{V_{cav}}$ reaches pattern specific limit $\sigma_{V_{b}}$ around $10^3$, so $S/N$ needs to larger than $10^3$ to minimize cavity voltage jitters. 

In Fig.~\ref{fig:patts-vs-SN} (b) and (c), we see $\sigma_{P_{amp}}$ and average ${P_{amp}}$ are sensitive to noise than filling pattern. To minimize power consumption $P_{amp}$ around to 11.15~kW, the $S/N$ has to be larger than $10^4$. Two patterns has similar amplifier power fluctuations $\sigma_{P_{amp}}$ up to $S/N$ = $10^{5}$. Beyond this point, $\sigma_{P_{amp}}$ reach filling pattern specific floors. 

The analytical model underestimates $P_{amp}$ as shown in Fig.~\ref{fig:patts-vs-SN} (b) at high noise. As the noise increase, the amplifier starts to have saturation. In this case, the proportional term can't provide sufficient power. As the power shortage build up, the integral term will start to make correction and add power the cavity. The simulation can model the proper PI controller and have integral term. But the analytical doesn't have the integral term and thus can't include the power from integral term. This will cause analytical model to fail at very high noise levels and accounts for the difference between the analytic model and simulation.

The typical $S/N$ range for a real LLRF system is around $10^{3} - 10^{6}$. In the figures, we cover a very wide range of $S/N$, including values which far exceed the realistic range of values. The reason for this is to allow us to explore the behaviour of the RF and LLRF system in the limit of ultra-low noise, which allows us to study features that are not visible at realisable values of $S/N$, such as the pattern-dependent noise floor in Fig.~\ref{fig:patts-vs-SN} (c).

%In Fig.~\ref{fig:patts-vs-SN} (b), average ${P_{amp}}$ agrees well with 

%Our theoretical model doesn't fit simulation very well, so introduced a filling pattern specific fudge factor $\alpha_{b}$ to the Eq.~\ref{eq:Amod12} and we get
%\begin{align}
%\sigma_{V_{cav}} &= \sqrt{ \langle V_{b}^2 \rangle + \alpha_{b}^2\langle V_{n}^2 \rangle },  
%    \label{eq:BLSim1}
%\end{align}
%\noindent where $\alpha_{b}$ is given by
%\begin{align}
%    \alpha_{b} = \frac{\sigma_{V_{b}^{\frac{1}{2}}}}{7}.   
%    \label{eq:BLSim1}
%\end{align}
%\noindent The fit results with and without $\alpha_{b}$ is given the Fig.~\ref{fig:patts-vs-SN} (a). $\alpha_{b}$ tunes down the effect of noise on $\sigma_{V_{cav}}$. We think this damping effect is because of the LLRF system and PI controller (proportional integral). The theoretical model doesn't include the PI controller, which uses a control loop feedback mechanism. We will further investigate this is issue in our future work.
\subsubsection{Cavity voltage}
\begin{figure}
	\begin{tabular}{c}
		{\scalebox{0.5} [0.5]{\includegraphics{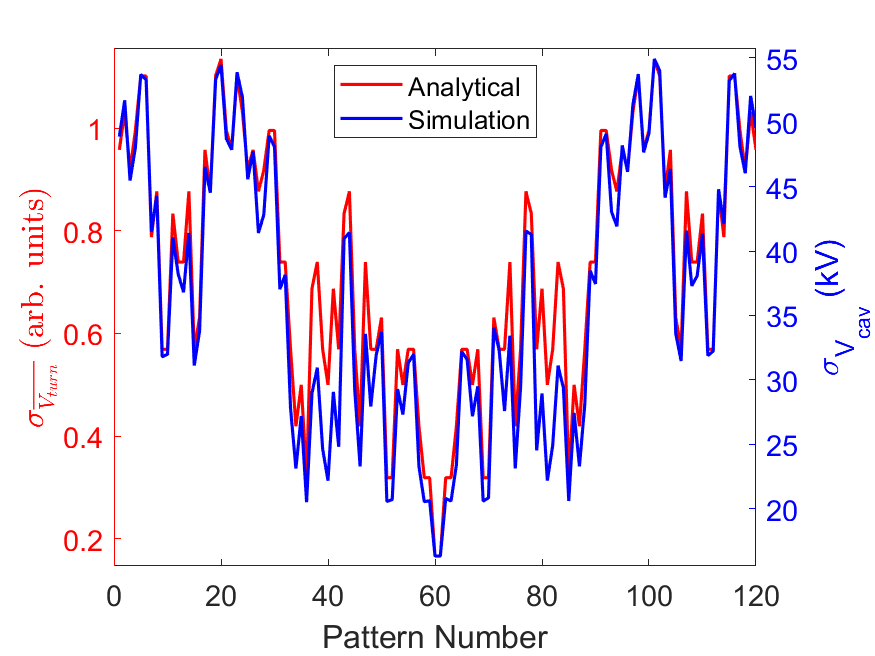}}} \\
		(a) \\
		{\scalebox{0.5} [0.5]{\includegraphics{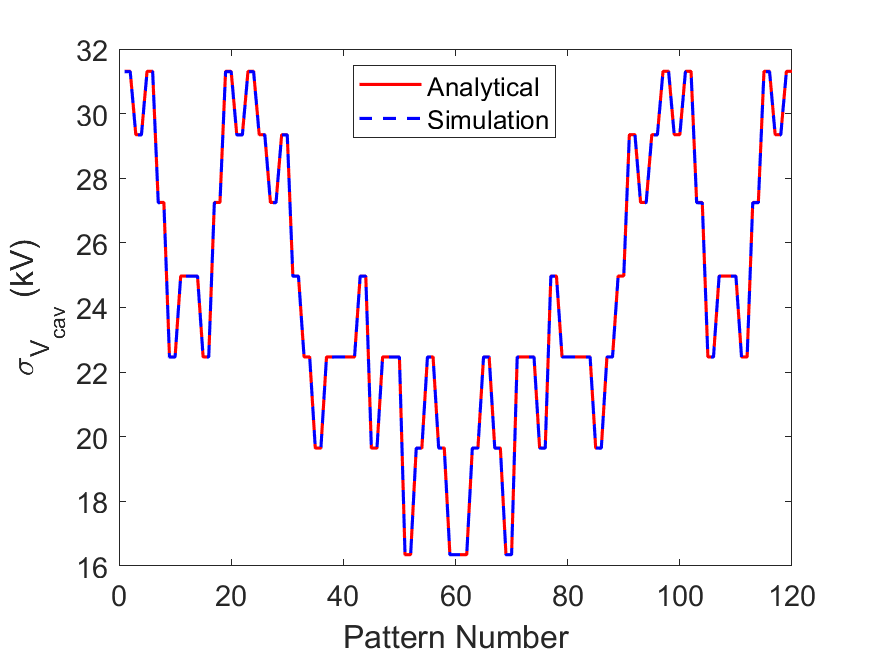}}}\\
		(b) \\
	\end{tabular}
	\caption{Simulated $\sigma_{V_{cav}}$ of 120 patterns with (a) static and (b) dynamic set-points compared to prediction. The $S/N$ was set to $1\times10^{12}$ to turn off the noise.}
	\label{fig:sim-sigVcav}
\end{figure}

The cavity voltages jitters $\sigma_{V_{cav}}$ of all 120 filling patterns are shown in Fig.~\ref{fig:sim-sigVcav}. We see that $\sigma_{V_{cav}}$ is different when different set-points are used. The dynamic set-point is better because it gives smaller cavity  voltage jitters. The  filling patterns No. 60 (pattern [1~4~3~6~5~2]) and 61 (pattern [1~4~5~2~3~6]) are optimum for both set-points. There are other patterns [1~4~2~5~3~6], [1~4~2~5~6~3], [1~4~3~6~2~5], [1~4~5~2~6~3], [1~4~6~3~2~5], and [1~4~6~3~5~2] are optimal only for dynamic set-point. This indicates that depending on the set-point type, the Figure Of Merit (FOM) to estimate $\sigma_{V_{cav}}$ is different. For static set-point, the FOM can be given as 

\begin{equation}
    \begin{split}
    \sigma_{V_{cav}} = \sigma_{\overline{V_{turns}}} = \sqrt{\frac{1}{N_{t}} \sum_{i=1}^{i=N_{t}}  (\bar{V}_{i})^2 },
    \label{eq:FOM-stat}
    \end{split}
\end{equation}

\noindent with ${\bar{V}_{i}}$ being the average voltage of $i^{th}$ turn, and $N_{t}$ being number of turns. In this case, we averaging voltage over one turn and get ${\bar{V}_{i}}$ first, then calculating the RMS of these $N_{t}$ turns. As shown in Fig.~\ref{fig:sim-sigVcav} (a), the FOM roughly overlaps with simulation. Although, the FOM doesn't predict jitters exactly, but it can find optimal pattern quickly without simulations. For dynamic set-point, the FOM is Eq.~\ref{eq:Amod7}. The theoretical prediction matches simulation results exactly for $S/N = 1\times10^{12}$ as shown in Fig.~\ref{fig:sim-sigVcav} (b). 

We see the dynamic set-point give smaller jitters. The patterns [1~4~3~6~5~2] and [1~4~5~2~3~6] (pattern number 60 and 61) are optimal in both set-points. Optimal pattern has 2$-$3 times less cavity voltage jitters than worst patterns.

\subsubsection{Amplifier power results}

\begin{figure}
	\begin{tabular}{c}
		{\scalebox{0.5} [0.5]{\includegraphics{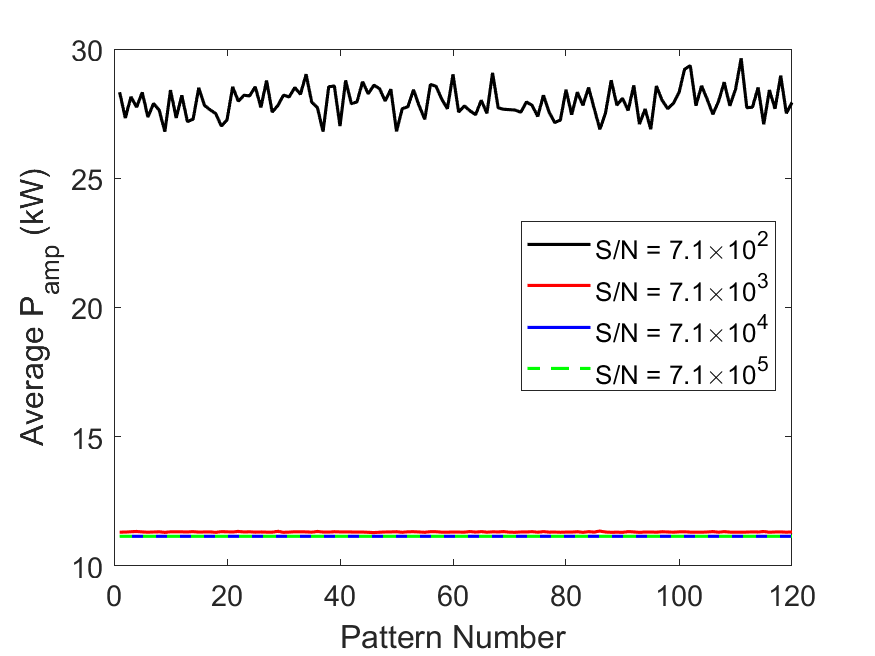}}} \\
		%{\scalebox{0.55} [0.55]{\includegraphics{3MeanPkly_SPDynamic_VcavKlyAmpScan2}}}\\
	\end{tabular}
	\caption{Average amplifier power $P_{amp}$ of 6-turn ERL patterns at different $S/N$.}
	\label{fig:ave-Pamp}
\end{figure}

The required average amplifier powers $P_{amp}$ for different patterns and different $S/N$ are given in Fig.~\ref{fig:ave-Pamp}. We see that the average $P_{amp}$ is reduced from 28~kW to 11.13~kW, when the $S/N$ increased from $7.1\times10^3$ to $7.1\times10^^4$. When $S/N$ reduced further, the $P_{amp}$ is reduced to minimum of 11.147~kW, which is the resistive power loss. This shows that ERLs can be operated with very low power, when $S/N$ is sufficiently high. 

\subsection{Property of optimal patterns}
\begin{figure}
	\begin{tabular}{c}
		{\scalebox{0.5} [0.5]{\includegraphics{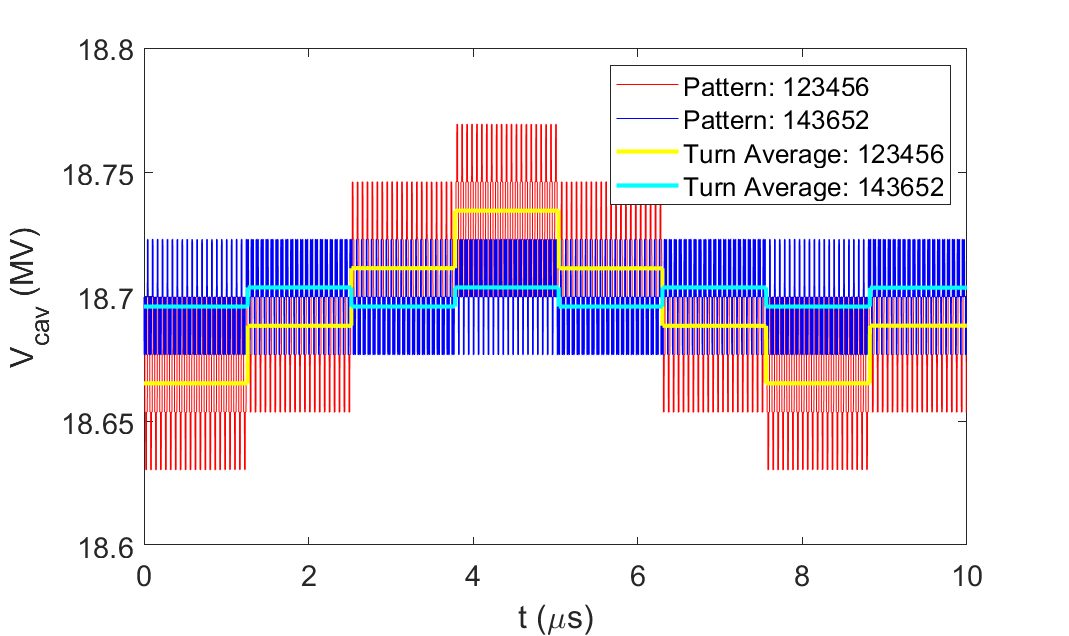}}} \\
		(a) \\
		{\scalebox{0.5} [0.5]{\includegraphics{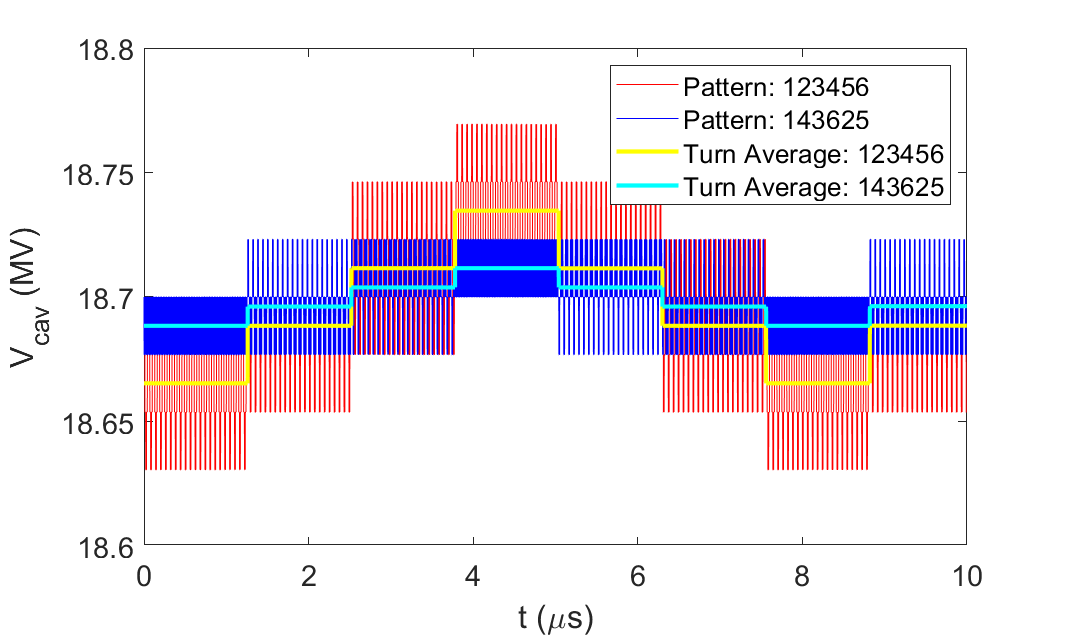}}}\\
		(b) \\		
	\end{tabular}
	\caption{Comparison of $V_{cav}$ and turn average of $V_{cav}$ of different patterns. (a) ASPO and non-optimal pattern. (b) DSPO and non-optimal pattern.}
	\label{fig:Vcavpatts}
\end{figure}

In Fig.~\ref{fig:Vcavpatts}, we compared cavity voltage of optimal and non-optimal patterns, indicated by blue and red lines respectively. In sub-figure (a), voltage of optimal pattern [1~4~3~6~5~2] fluctuates less than $\pm0.024$~MV range of 18.7~MV, while non-optimal pattern [1~2~3~4~5~6] has 3 times larger fluctuation. We see similar 3-up-3-down and up-down fluctuations as in Fig.~\ref{fig:6Turn_BL_patterns}, but here we have 20 bunch packets, so these fluctuations are repeated 20 times in each turn. Revolution times is about 1.2~$\mu$s, so every 1.2~$\mu$s turn changes. 

The optimum filling patterns [1~4~3~6~5~2] and [1~4~5~2~3~6] (pattern number 60 and 61) and their associated beam loading patterns are given in Table~\ref{tab:bl_patt}. We observe their two consecutive bits are in either up-down (10) or down-up (01) pairs. Such combinations limit cumulative sum of beam loading pattern to a range of [$-1$, 1], and thus minimizes jitters. We also see 1 pair flips (``1" and ``0" switch positions) per turn. The change from ``0" to ``1" (acceleration to deceleration) happens in 3rd to 4th turn transition and the change from ``1" to ``0" is the new bunch replacing the extracted bunch. Therefore, in optimal patterns, consecutive pairs are made up by bunches that are 3 turns apart like [1~4], [2~5], and [3~6].

Patterns [1~4~2~5~3~6], [1~4~2~5~6~3], [1~4~3~6~2~5], [1~4~5~2~6~3], [1~4~6~3~2~5], and [1~4~6~3~5~2] also have above motioned properties of optimal patterns. However, they are only optimal for dynamic set-point and not for static set-point. Therefore, these 6 patterns are Dynamic Set-Point Optimal (DSPO) patterns, while [1~4~3~6~5~2] and [1~4~5~2~3~6] are All Set-Point Optimal (ASPO) patterns. Of course, a ASPO pattern is a DSPO pattern by definition. The difference between the ASPO pattern [1~4~3~6~5~2] and DSPO pattern [1~4~3~6~2~5] is shown in Fig.~\ref{fig:Vcavpatts}. Both patterns have same fluctuation range, but the turn average of the DSPO is larger in the 1st, 4th, and 7th turns. So, $\sigma_{\overline{V_{turn}}}$ of pattern DSPO is larger, which makes it non-optimal for static set-points according to Eq.~\ref{eq:FOM-stat}. 

\subsection{Off-crest beam loading}
\begin{figure}
		{\scalebox{0.3} [0.3]{\includegraphics{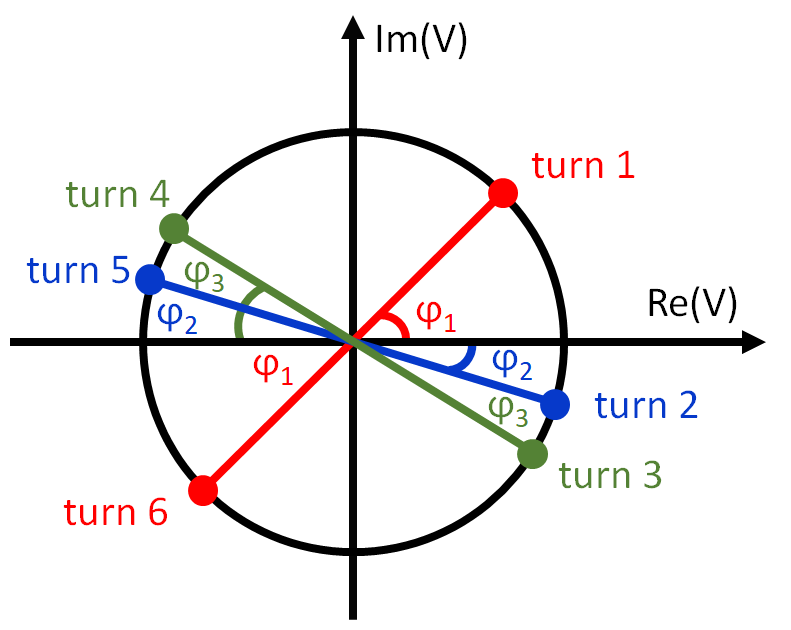}}}\\
	\caption{Definition of off-set angels in off-crest beamloading.}
	\label{fig:off-crest-bl}
\end{figure}

% to prevent building up of non-linear effects~\cite{PeterERLOffcrestBL} and bunch compression for FEL applications.
So far, we have studied the effects of beam loading for on-crest phases. In applications such as FELs, bunches must be compressed during acceleration to achieve high peak current, then stretched and energy compressed on deceleration to eliminate adiabatic energy spread growth. Beams must therefore pass through the RF system off crest~\cite{AChaoHandBook, Neil_PRL_ERL_2000}. In recirculating ERLs, we want to minimize the net beam loading of a packet, so the in-phase (I) and quadrature phase (Q) components of the beam loading of a packet should sum to approximately zero, i.e. the vector sum of the voltage changes sums to zero for the bunch packet. By doing so, the amplitude and phase of the cavity voltage changes minimally after a packet. This implies that the phase and amplitude perturbations from beam loading cancel out over a bunch packet, as shown in Fig.~\ref{fig:off-crest-bl}. Here, by ``mirror turns" we meant turns that has same energy but the bunch phase is offset by $\pi$ radians. In 6-turn ERLs, turn 1 and 6, 2 and 5, and 3 and 4 are mirror turns. Mirror bunches have same energy and off-set angles as shown in Fig.~\ref{fig:off-crest-bl}, so their vector sum is zero. In Fig.~\ref{fig:off-crest-bl}, $\phi_{1}$ is the off phase angle of 1st and 6th turns; $\phi_{2}$ is the off phase angle of 2nd and 5th turns; $\phi_{3}$ is the off phase angle of 3rd and 4th turns.

\subsubsection{Phase angle jitters}
\begin{figure}
		%{\scalebox{0.5} [0.5]{\includegraphics{2Phasecav_NoSequencePerserving_SetPointDynamic_SN1e12}}}\\    
		{\scalebox{0.5} [0.5]{\includegraphics{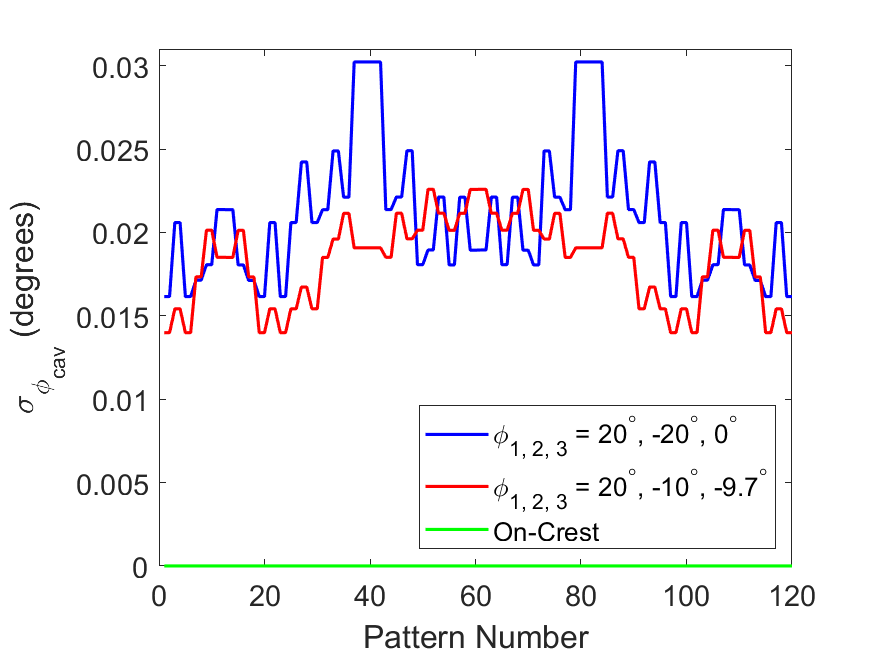}}}\\   		
	\caption{Cavity voltage phase jitters of off-crest beam loading for 120 patterns for 6-turn ERL.}
	\label{fig:off-crest-phase}
\end{figure}

We have estimated off-crest cavity voltage phase fluctuation for 120 patterns of the 6-turn ERL and results are given in Fig.~\ref{fig:off-crest-phase}. The $S/N$ was set to $10^{12}$ to turn off the noise. We have simulated two sets of off-set angles $\phi_{1, 2, 3} = 20^{\circ}, -20^{\circ}, 0^{\circ}$ and $\phi_{1, 2, 3} = 20^{\circ}, -10^{\circ}, -9.7^{\circ}$. We see that: (1) phase jitters is pattern dependent; (2) phase jitters is off-phase angle dependent; (3) in the worst case scenario, the RMS cavity phase jitters is less than $0.03^{\circ}$, even at fairly large off-set angles. (4) the jitters in the on-crest case is negligible. 

For the two ASPO patterns (pattern number 60 and 61), the first off-set angles $\phi_{1, 2, 3} = 20^{\circ}, -20^{\circ}, 0^{\circ}$ has smaller jitters of $0.019^{\circ}$. The $\sigma_{\phi_{cav}}$ pattern is approximately up-side down of $\sigma_{V_{cav}}$, as can be seen from Figs~\ref{fig:off-crest-phase} and \ref{fig:off-crest-Vcav} (a). This is more obvious for $\phi_{1, 2, 3} = 20^{\circ}, -10^{\circ}, -9.7^{\circ}$ angle sets. This indicates if a pattern has larger amplitude jitters, then it tends to have smaller phase jitters, and visa versa. 

\subsubsection{Cavity voltage and amplifier power jitters}
\begin{figure}
	\begin{tabular}{c}
		{\scalebox{0.5} [0.5]{\includegraphics{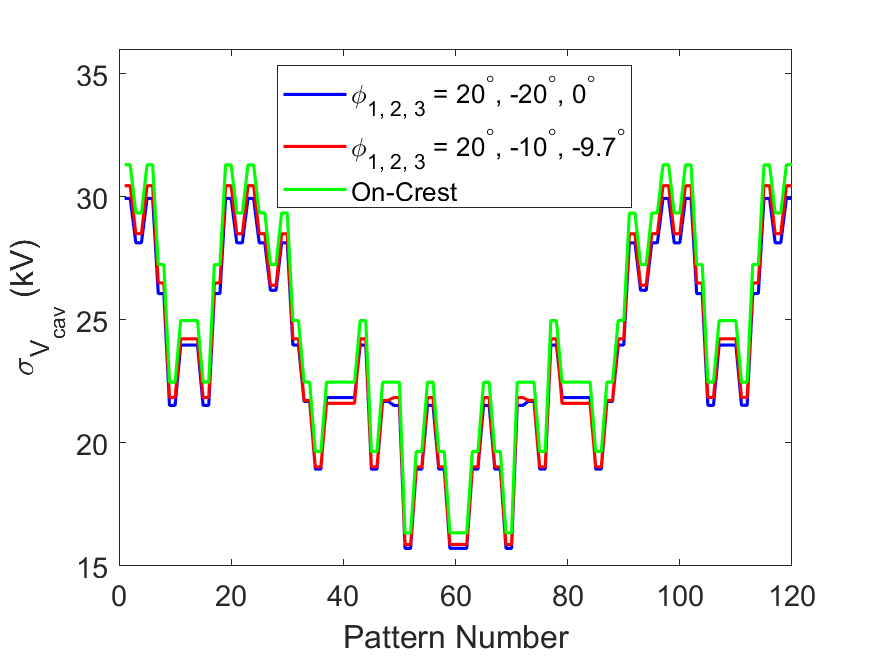}}} \\
		%{\scalebox{0.5} [0.5]{\includegraphics{1Vcav_NoSequencePerserving_SetPointDynamic_SN1e12}}} \\
		(a) \\
		{\scalebox{0.5} [0.5]{\includegraphics{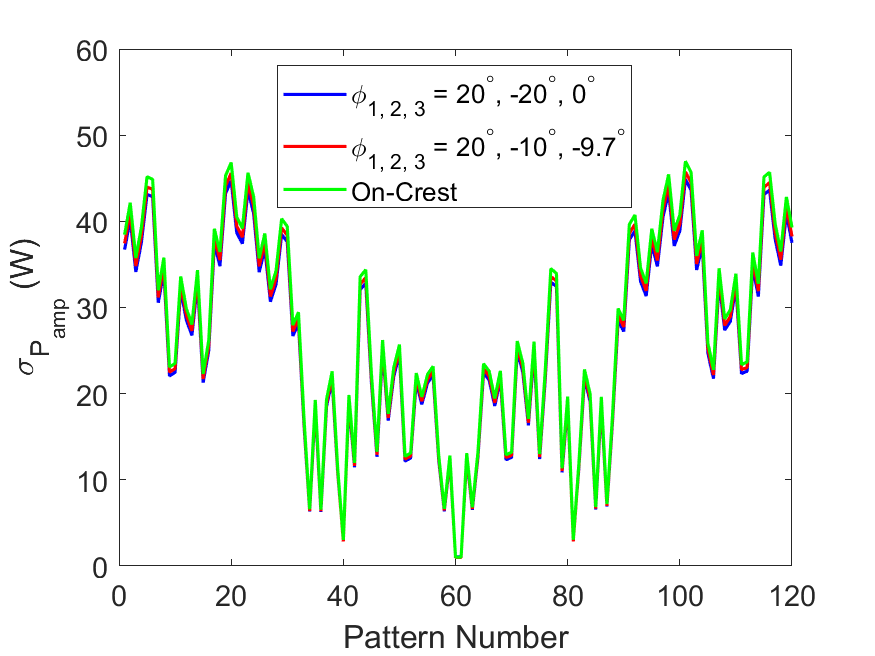}}}\\
		%{\scalebox{0.5} [0.5]{\includegraphics{3RMSPkly_NoSequencePerserving_SetPointDynamic_SN1e12}}}\\
		(b) \\	
%		{\scalebox{0.5} [0.5]{\includegraphics{4MeanPkly_SequencePerserving_SetPointDynamic_SN1e12}}}\\
%		(c) \\
	\end{tabular}
	\caption{Cavity voltage fluctuation (a) and amplifier power amplitude fluctuation (b) of off-crest beam loading for 120 patterns for 6-turn ERL. The S/N is 10$^{12}$.}
	\label{fig:off-crest-Vcav}
\end{figure}

We have also estimated cavity voltage and amplifier power jitters and results are given in Fig.~\ref{fig:off-crest-Vcav}. The difference in on- and off-crest cases are insignificant. The average amplifier power is the same as on-crest case, which is about 11.15~kW for all filling patterns. 

\subsection{Bunch charge jitter}

Bunch charge modulations for a recirculating ERL introduces a unique source of noise that is unlike other sources we have considered thus far in this article. An error on bunch charge persists over all turns in the ERL before the beam is dumped. As a result, the noise spectrum from charge modulation is significantly narrower than the white noise we have assumed for other noise sources. For the 6-turn ERL we consider in this paper, the effective noise spectrum for the bunch charge jitter is peaked at approximately 140 kHz, and therefore it is within the closed-loop bandwidth of 2.5 MHz for the LLRF controller. For small bunch charge errors, the LLRF system is easily able to correct the error, whereas for larger values, it will struggle and the charge jitter becomes the dominant noise source.

\begin{figure*}
	\begin{subfigure}[b]{0.32\textwidth}
		\includegraphics[width=\textwidth]{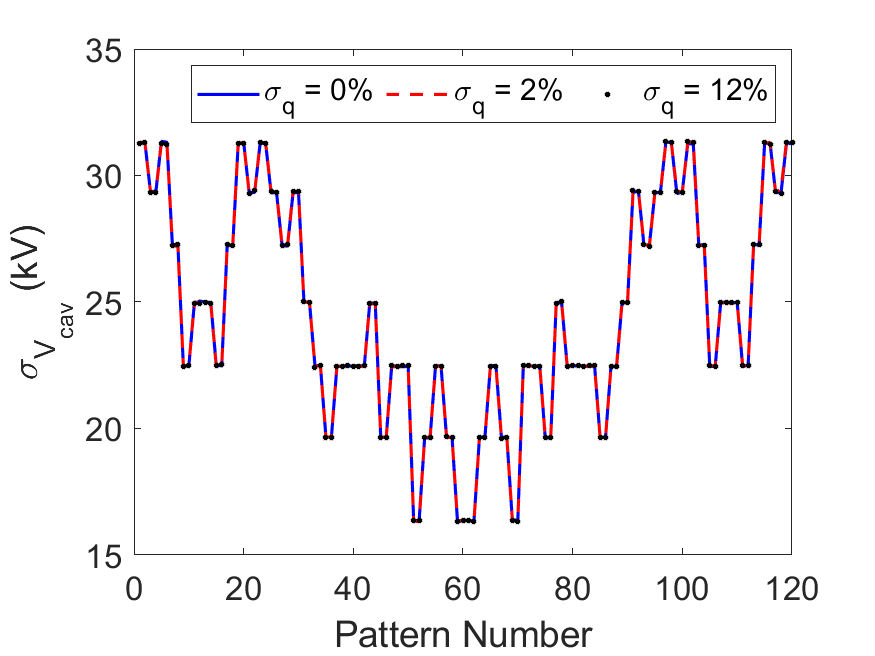}
		\caption{Dynamic set-point}
	\end{subfigure}
	\begin{subfigure}[b]{0.32\textwidth}
		\includegraphics[width=\textwidth]{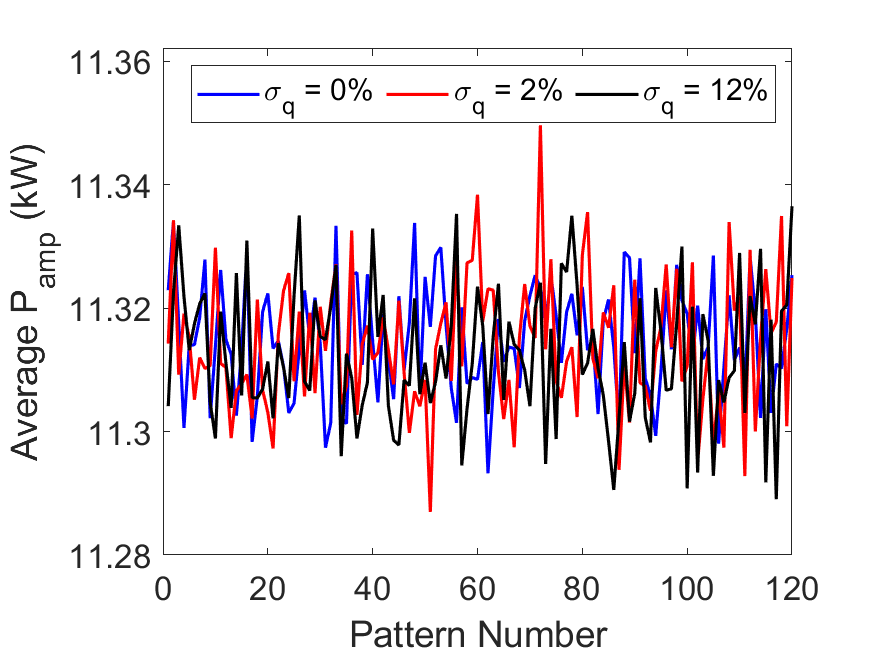}
		\caption{Dynamic set-point}
	\end{subfigure}
	\begin{subfigure}[b]{0.32\textwidth}
		\includegraphics[width=\textwidth]{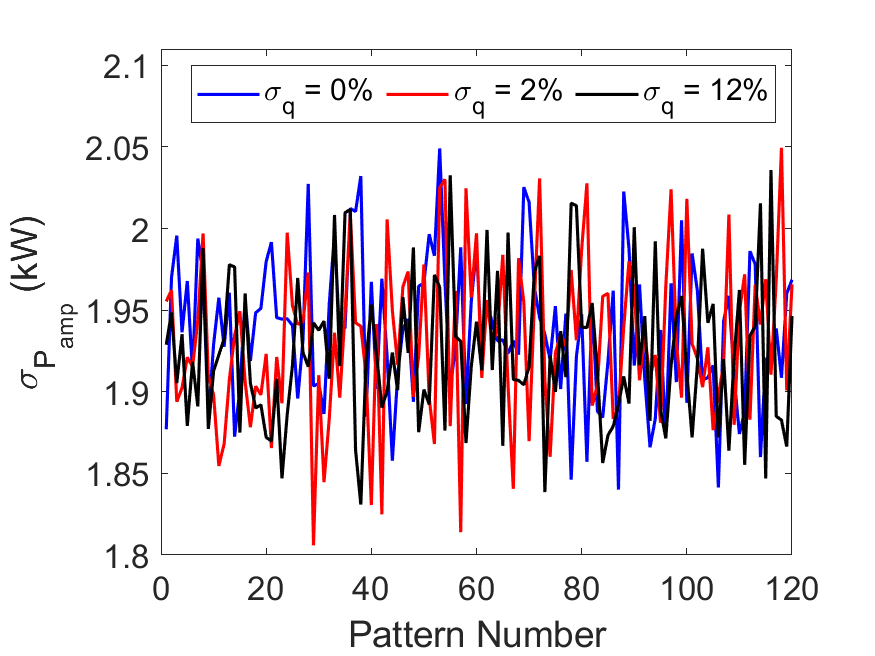}
		\caption{Dynamic set-point}
	\end{subfigure}
	\begin{subfigure}[b]{0.32\textwidth}
		\includegraphics[width=\textwidth]{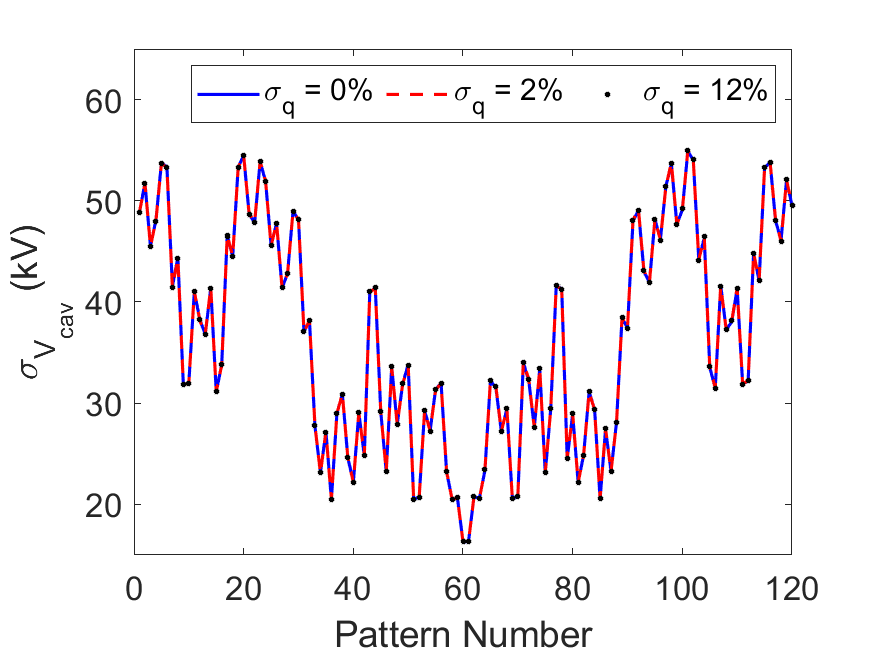}
		\caption{Static set-point}
	\end{subfigure}
	\begin{subfigure}[b]{0.32\textwidth}
	\includegraphics[width=\textwidth]{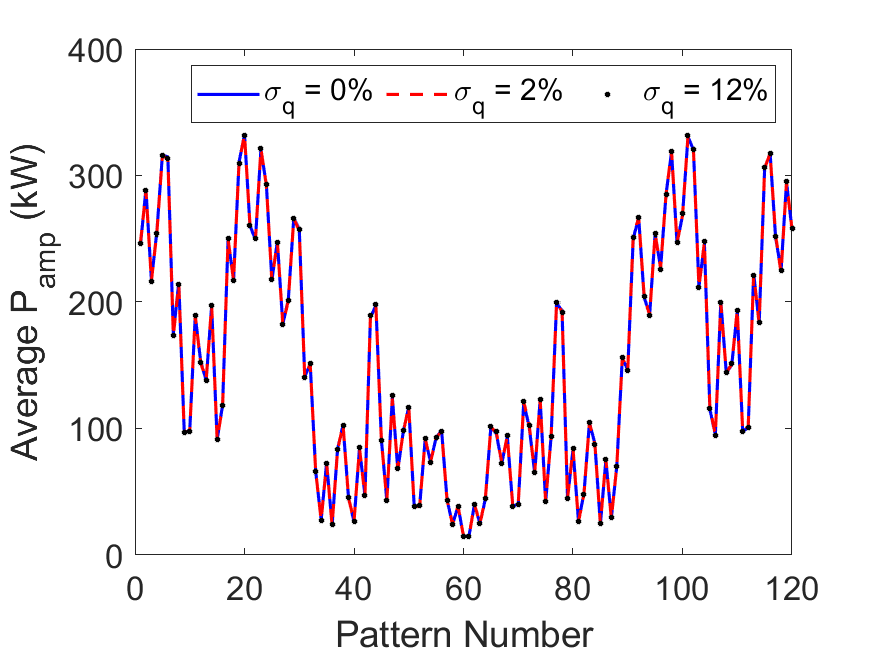}
	\caption{Static set-point}
	\end{subfigure}
	\begin{subfigure}[b]{0.32\textwidth}
	\includegraphics[width=\textwidth]{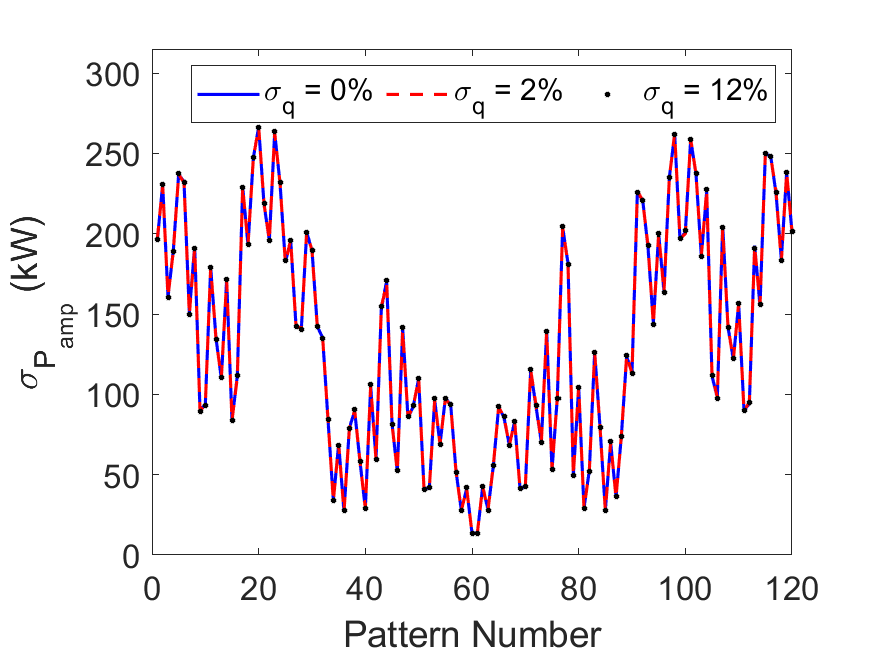}
	\caption{Static set-point}
	\end{subfigure}
	\caption{Bunch charge jitter simulation results with RMS bunch jitters of 0, 2$\%$, and 12$\%$. (a), (b), and (c) are results with dynamic set-point; (d), (e), and (f) are results with static set-point.}
	\label{fig:charge-jitter}
\end{figure*}

We performed beam loading simulations to investigate effect of bunch charge jitter on the cavity voltage and amplifier power. The jitter was assumed to be Gaussian. RMS bunch charge jitters with 2$\%$ and 12$\%$ were simulated. Simulations were carried out for 120 filling patterns with the S/N = 7100, bunches launched on crest, and both set-points. The results are given in Fig.~\ref{fig:charge-jitter} for RMS cavity voltages in sub-figures (a) and (d), for average amplifier powers in (b) and (e), and RMS amplifier jitters in (c) and (f). The sub-figures (a), (b), and (c) are results for dynamic set-points and (d), (e), and (f) are for static set-points. We see charge jitters does not increase cavity voltage jitters for both static and dynamic set-points, even when $\sigma_{q}~=$ 12$\%$. We see the filling pattern and other noises are dominant over charge jitter noise. 

%The average amplifier power and power jitters are increased in static set-point case, but not in the dynamic set-point. When the RMS charge jitter is 2$\%$, its effect on cavity voltage and amplifier are insignificant for both set-points. However, when it is 12$\%$, the charge jitter over takes pattern specific beam loading jitter to become dominant factor and all the patters have similar jitters. Therefore, it is important to keep bunch charge jitter to less than 2$\%$ to lower cavity voltage fluctuation and average amplifier power.

%The effect of bunch charge jitter on the beam energy is a secondary effect as it acts through cavity voltage. When $\sigma_{q}~=$ 2$\%$, $\sigma_{V_{cav}}$ is about 20 kV, of which 1-2 kV by bunch charge jitter and rest by beam loading without bunch charge jitter. So, bunch charge jitter is one order magnitude less than beam loading jitter, in terms of its effect on $\sigma_{V_{cav}}$. Therefore, its effect on the beam energy should also be 1 order magnitude less than beam loading jitter. Only when $\sigma_{q}~=$ 12$\%$, they become comparable. Most injectors can achieve less few percents charge jitters. Therefore, the effect of charge jitter on the cavity voltage, amplifier and beam energy are less significant than beam loading without charge jitter.

\subsection{Energy modulation}

It is possible that disturbances, such as charge jitter, beam loading, or other noise or jitter sources, may result in an energy modulation on the accelerating or decelerating beam. The stored energy in the cavity is given in the Eq.~\ref{eq:ustored}. Therefore, the change in energy of the cavity when a beam passes through is equal to minus the energy change of the particle bunch as it passes through the cavity ($q_{bunch}V_{cav}e^{j\phi}$), where $\phi$ is the RF phase at which the bunch passes through the cavity:

\begin{equation}
    \delta{U_{stored}}=\frac{\left(V_{cav}+\delta{V}\right)^{2}-V_{cav}^{2}}{\omega\left(\frac{R}{Q}\right)}=q_{bunch}V_{cav}e^{j\phi}
    \label{eq:energymod2}
\end{equation}

Usually, Eq.~\ref{eq:energymod2} is simplified to a linear approximation by assuming that the change in cavity voltage is small compared to the cavity voltage, in which case, we obtain $\delta{V}=\frac{q_{bunch}\omega}{2}\left(\frac{R}{Q}\right)e^{j\phi}$, which is independent of the cavity voltage, and small modulations on the cavity voltage do not lead to an energy modulation on the bunches. However, if we don't approximate Eq.~\ref{eq:energymod2}, we get that the change in cavity voltage due to beam loading is:

\begin{equation}
\begin{aligned}
    \delta{V}=-V_{cav}\left[1+\sqrt{1-\frac{q_{bunch}\omega}{V_{cav}}\left(\frac{R}{Q}\right)}e^{j\phi}\right] \\
    \approx \frac{q_{bunch}\omega}{2}\left(\frac{R}{Q}\right)e^{j\phi}\left[1+\frac{q_{bunch}\omega}{4V_{cav}}\left(\frac{R}{Q}\right)e^{j\phi}+\cdots\right]
    \label{eq:energymod3}
\end{aligned}
\end{equation}

The second term in Eq.~\ref{eq:energymod3} does result in an energy modulation, and in fact it is the dominant term for causing an energy modulation. If we use the values from Table~\ref{tab:sim-setup}, we find that the second term in Eq.~\ref{eq:energymod3} is approximately 0.06\% of the magnitude of the first term. Therefore, the resultant energy modulation caused by beam loading in our hypothetical recirculating ERL is negligible, hence the energy modulation due to effects such as charge jitter will be even smaller and for most scenarios it can be neglected. However, if we operate at very high frequency ($\sim$THz), very high bunch charge (which would exceed the threshold current for an ERL), or the cavity operates at very low voltages ($<$ kV) then the higher order terms in Eq.~\ref{eq:energymod3} become significant. This would also mean that the machine is operating in a non-linear regime, which would not be beneficial.

%The cavity voltage is 18.7~MV and $\sigma_{V_{cav}}$ tens of kV, which is 3 orders of magnitude difference. 

%We see cavity voltage fluctuation for 2$\%$ full charge jitter is insignificant and for 2$\%$ only increases about 1 kV. This is one order of magnitude less than the jitters by other sources, which is about 15-30 kV, depending on the filling pattern. We intentionally chose huge charge jitter of 12$\%$ to estimate system behaviour at worst case scenario, but most injectors can achieve much less charge jitters (few percents). Therefore, the effect of charge jitter on the cavity voltage, amplifier and beam energy are insignificant for standard decent quality  injectors. 

\section{Sequence preserving scheme}
\label{section:4}
\begin{figure}
		{\scalebox{0.35} [0.35]{\includegraphics{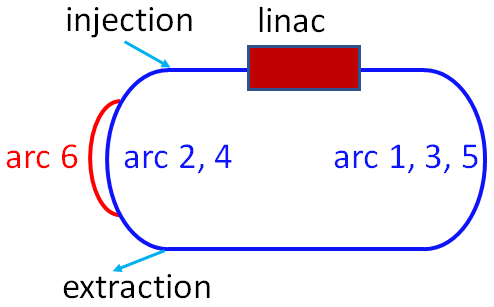}}} \\
	\caption{Topology with extra arc length for phase flip and/or delay.}
	\label{fig:8-toplogy}
\end{figure}

For a recirculating linac to be an ERL, there has to be an extra path length to delay the bunch by 180$^\circ$ phase to switch from accelerating mode to decelerating mode. By adjusting the delay length or by implementing more sophisticated arcs, topologies, and injection scheme, one can manipulate the bunch order or bunch spacing. The extra path length can be in the form of a longer arc~\cite{HIGH-CURRENT-ERL} or a chicane~\cite{CEBAF-ERL-EXP}. By introducing this additional path length, the topology changes from the ``0" topology of Fig.~\ref{fig:SimpleERL} to the ``8" topology of Fig.~\ref{fig:8-toplogy}. More complicated topologies can be achieved by setting all the arcs to different lengths~\cite{Douglas:IPAC2018-THPMK106, HZBERLCDR, Angal-Kalinin2018}. 

\begin{figure}
		{\scalebox{0.45} [0.45]{\includegraphics{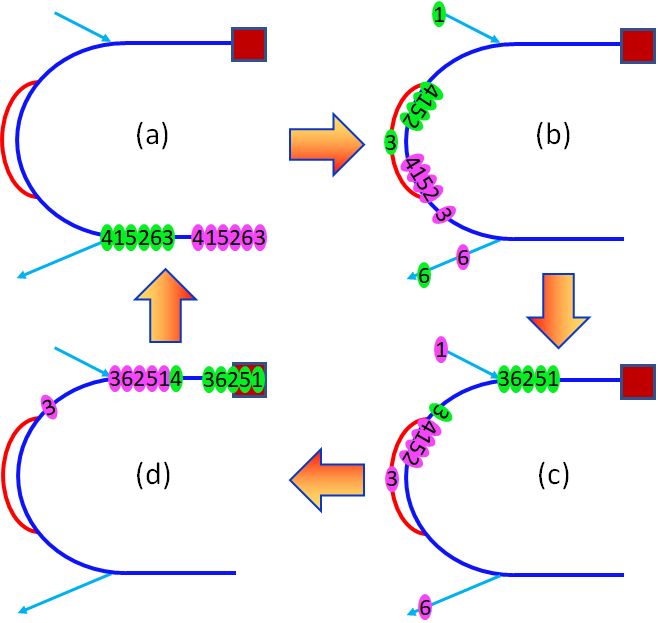}}} \\
	\caption{Topology with an extra arc 6 length to preserve $\{4~1~5~2~6~3\}$ bunch-turn number sequence. (a) Depiction of two bunch packets before entering the arcs. (b) Green bunch at 3$^{\text{rd}}$ turn gets delayed. Bunches at their 6$^{\text{th}}$ turn are extracted. (c) Green bunch at 3rd turn is delayed and joined pink packet. When the packet passes injection point, all the bunch numbers are incremented by 1. (d) A new bunch is injected into pink packet. New circulation starts with (a) again.}
	\label{fig:delay-scheme}
\end{figure}

Here we discuss ``8" topology as an example to show that it can maintain an `up-down-up-down' ([1~0~1~0~1~0]) beam loading pattern for all turns; which is preferable for cavity voltage and RF stability. It is achieved by utilising an injection and delay scheme shown the Fig.~\ref{fig:delay-scheme}. Such a scheme preserves $\{4~1~5~2~6~3\}$ bunch-turn number sequence and [1~0~1~0~1~0] beam loading pattern. Bunch-turn number sequence $\{4~1~5~2~6~3\}$ indicates the first bunch of bunch packet is at 4th turn, the second bunch is at 1st turn and so on. In SP schemes the new bunch is injected to the head of the packet and the bunch 3 of the earlier packet is delayed to join subsequent packet. In the previously described FIFO scheme, the new bunch is injected to the position of the dumped bunch and thus the bunch-turn number sequence changes turn-by-turn. 

Of course, one can maintain `up-down-up-down' patterns with more complicated topologies as well. The presented SP pattern is suitable for both simple or complicated topologies as it can maintain the favoured `up-down-up-down' beam loading pattern and there is no difference from the RF system perspective. For this SP scheme, the cavity voltage fluctuates within $\pm0.5$ normalized beam loading increment, which is half of the optimal FIFO patterns. However, FIFO patterns can achieve a higher density of bunch packets than SP patterns as it is necessary for SP patterns to have unoccupied intra-packet blocks to allow for the required manipulation of the bunch packet to maintain a constant beam loading pattern.

In ``8" topology of Fig.~\ref{fig:8-toplogy}, all bunches go through the same arc, except for the bunch transitioning from accelerating to decelerating modes. The transitioning bunch goes through arc 6, which has extra length $\Delta{L}$ for delay. The length of delay can be given as 

\begin{equation}
\Delta L = nL_{packet} + mL_{block} + \frac{\lambda_{RF}}{2}.
    \label{eq:delayDL}
\end{equation}

\noindent with $n = 0, 1, 2, ...$, $m = 0, 1, 2, ...$, $L_{packet}$ being the length occupied by a bunch packet, $L_{block}$ being the length occupied by a intra-packet block, and $\lambda_{RF}$ being the wave length of RF cycle. When $m=n=0$, the bunch flips phase but remains in the same packet; which is the case of the simple recirculating FIFO scheme described in earlier sections. The beam line layout described in \cite{CEBAF-ERL-EXP} can be an example of this. When $m,n\neq0$, the bunches don't only flip phase, but also move to later blocks and packets. 

%An example for case (2) is given in Fig.~\ref{fig:delay-scheme} when $n = 0$ and $m = 1$. 

Note that sequence $\{4~1~5~2~6~3\}$ indicates the turn number of bunches and should not be confused with filling pattern [1~5~2~6~3~4], which describes filling order. Angal-Kalinin $et$ $al.$, proposed~\cite{Angal-Kalinin2018} a similar SP scheme as $\{4~1~5~2~6~3\}$ for the purpose of separating low energy bunches to minimize Beam-Breakup (BBU) instability~\cite{Hoffstaetter_2004}. BBU is a major limiting factor for the ERL beam current~\cite{Lou_Hoffstaetter_2019} and we will investigate it further in a future study.  

\section{Comparison of simulation results}
\label{section:5}

\begin{figure*}
  \begin{subfigure}[b]{0.45\textwidth}
    \includegraphics[width=\textwidth]{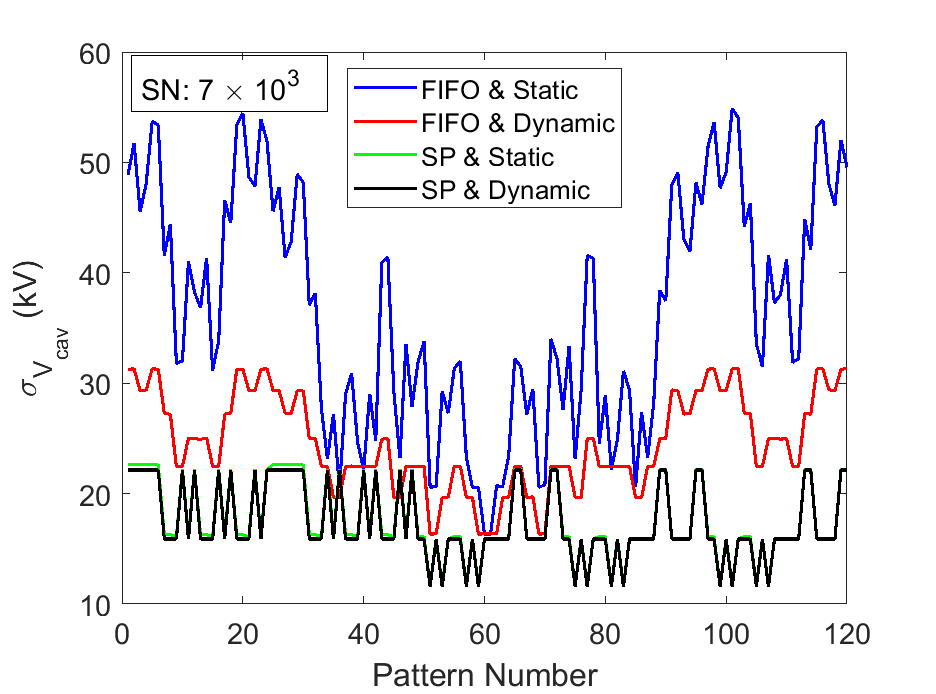}
    \caption{}
  \end{subfigure}
  \begin{subfigure}[b]{0.45\textwidth}
    \includegraphics[width=\textwidth]{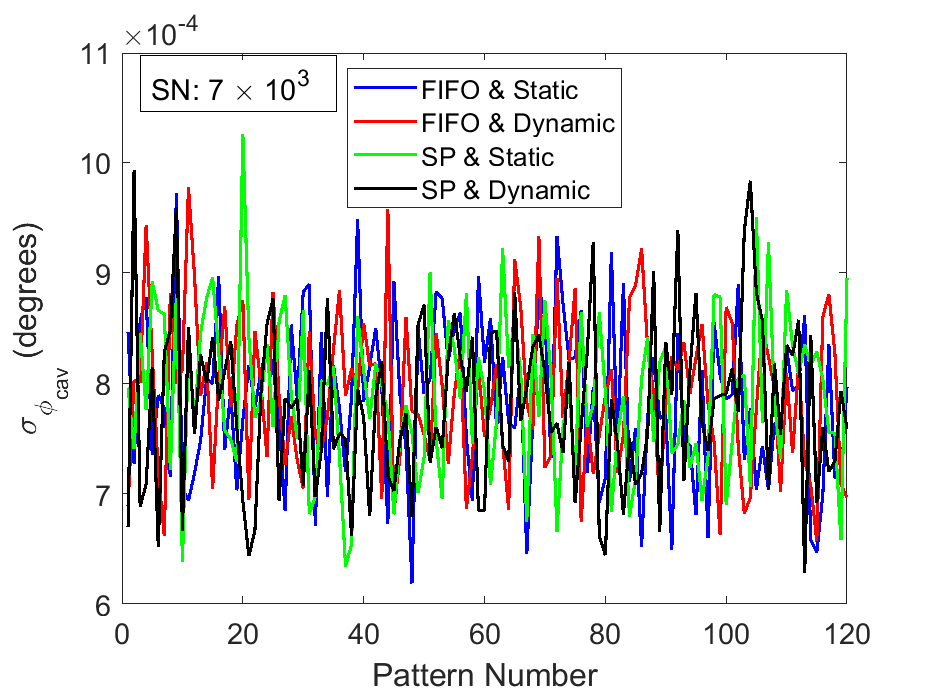}
    \caption{}
  \end{subfigure}
    \begin{subfigure}[b]{0.45\textwidth}
    \includegraphics[width=\textwidth]{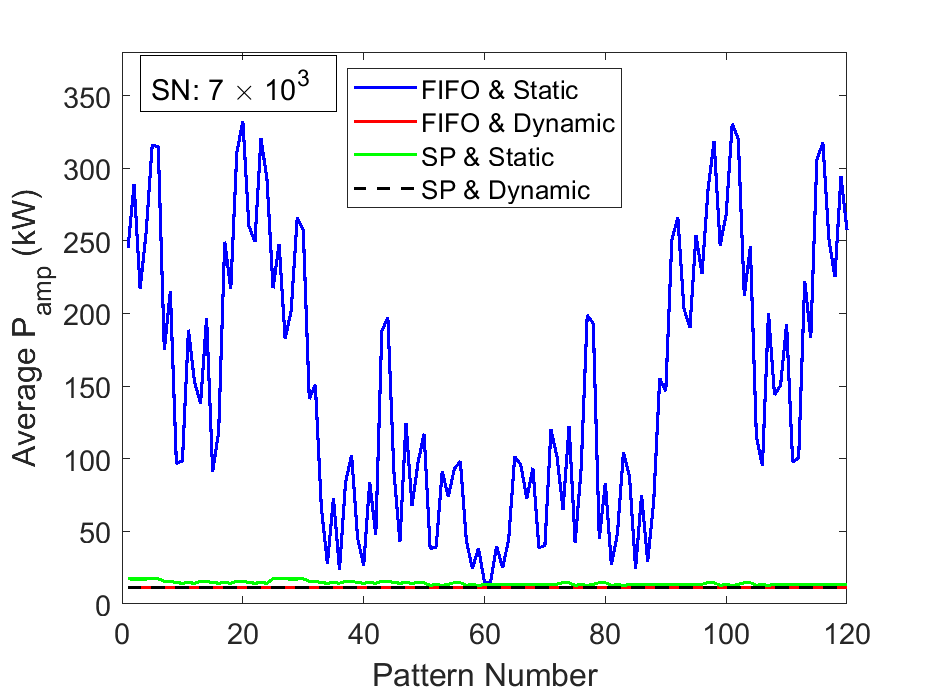}
    \caption{}
  \end{subfigure}
  \begin{subfigure}[b]{0.45\textwidth}
    \includegraphics[width=\textwidth]{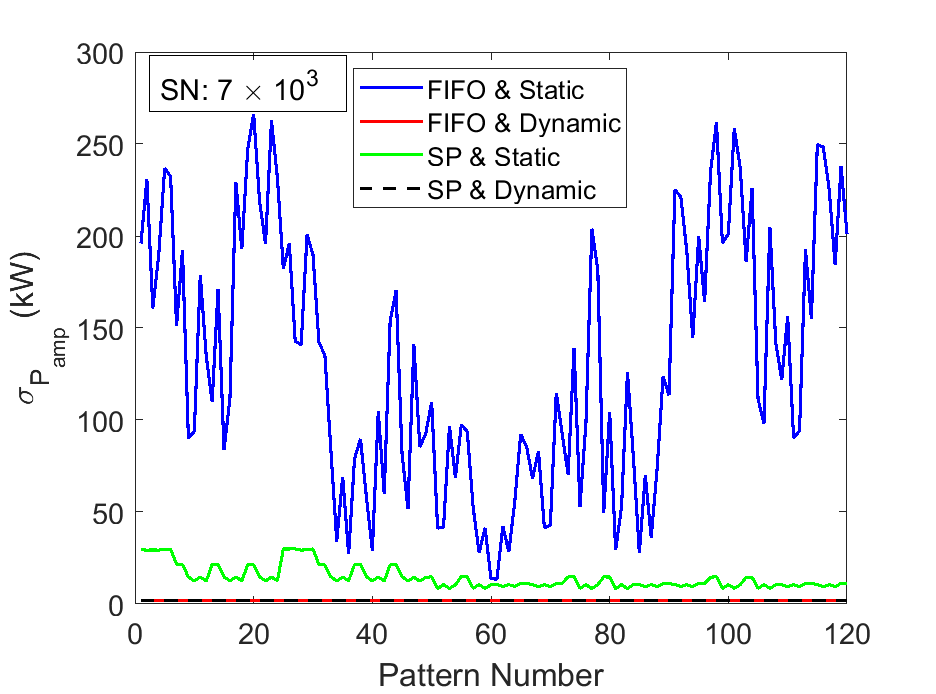}
    \caption{}
  \end{subfigure}
  	\caption{Comparison of SP and FIFO at dynamic set-point with on-crest beam loadings: (a) cavity voltage jitters; (b) cavity phase jitters; (c) average amplifier power; and (d) amplifier power jitters.}
  	\label{fig:22}
\end{figure*}

\begin{figure*}
	\begin{subfigure}[b]{0.49\textwidth}
		\includegraphics[width=\textwidth]{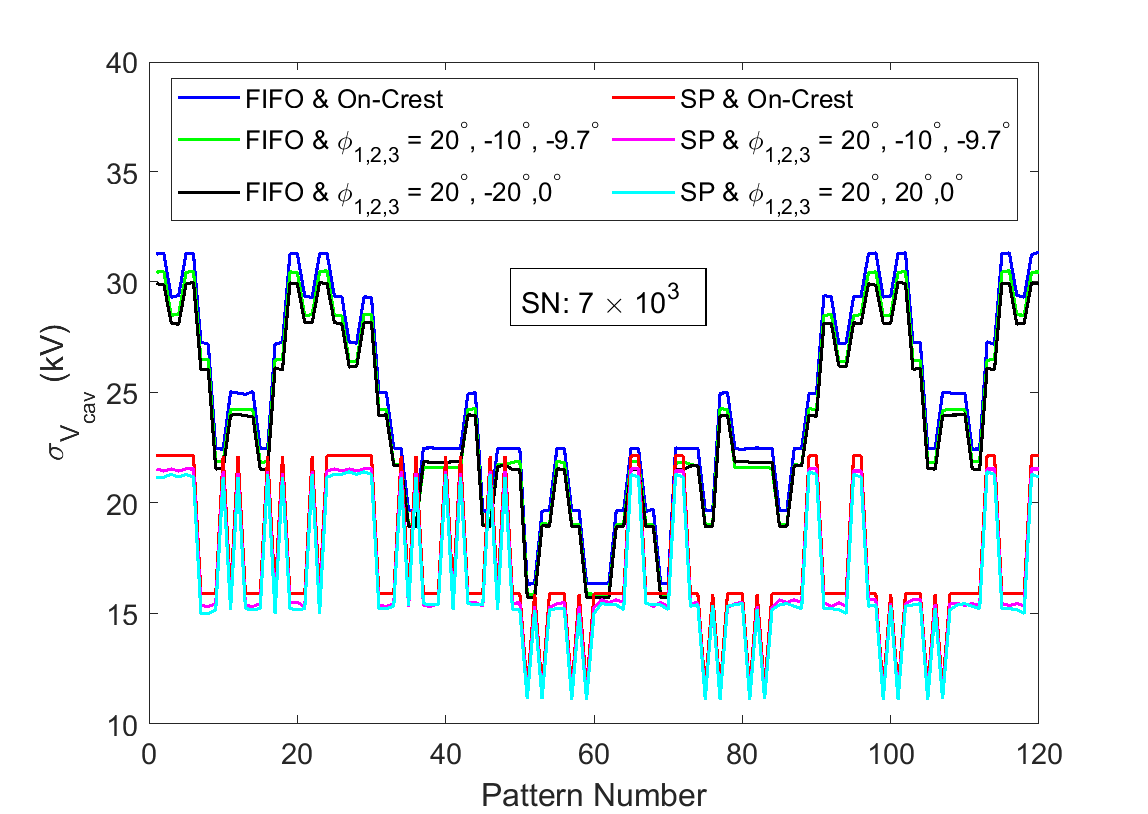}
		\caption{}
	\end{subfigure}
	\begin{subfigure}[b]{0.49\textwidth}
		\includegraphics[width=\textwidth]{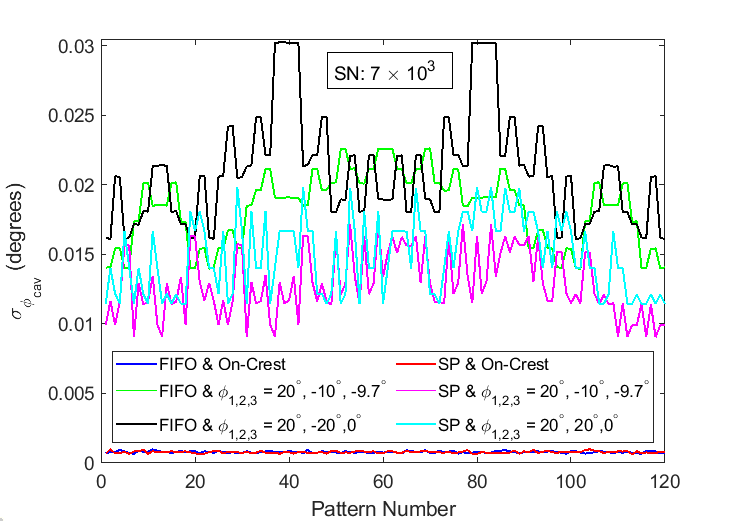}
		\caption{}
	\end{subfigure}
	\begin{subfigure}[b]{0.49\textwidth}
		\includegraphics[width=\textwidth]{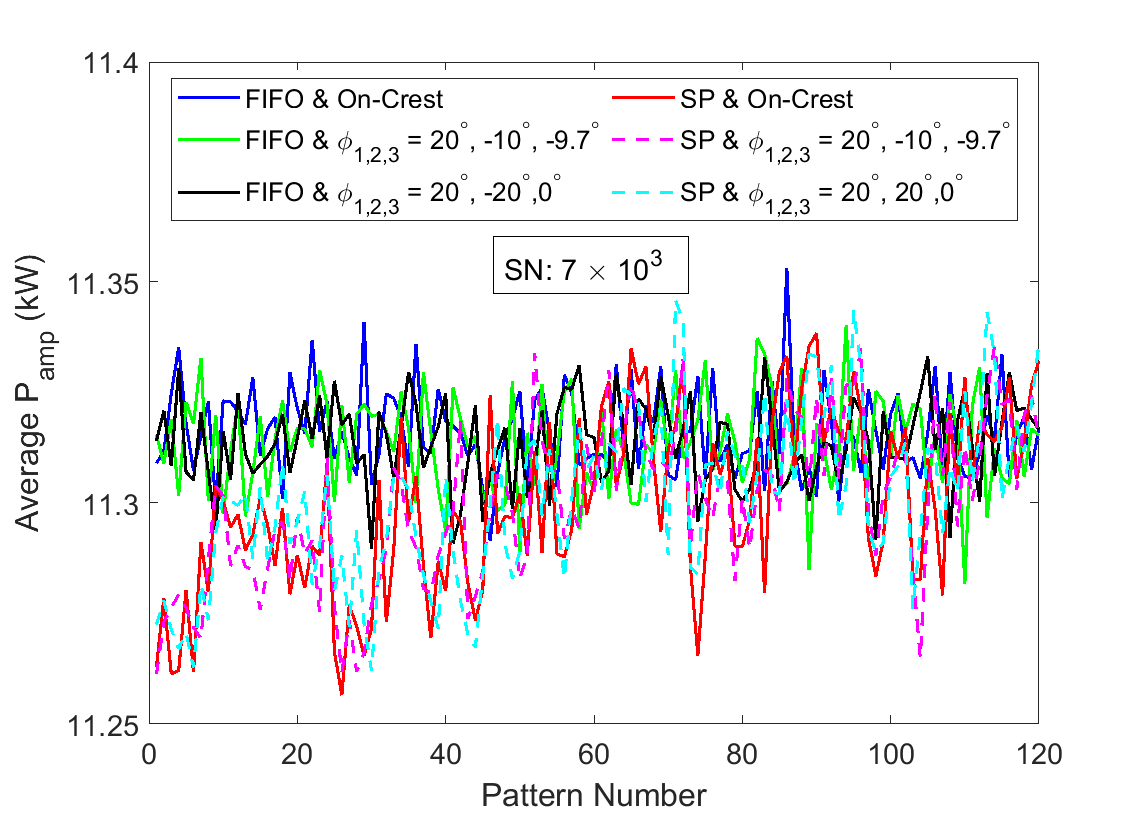}
		\caption{}
	\end{subfigure}
	\begin{subfigure}[b]{0.49\textwidth}
		\includegraphics[width=\textwidth]{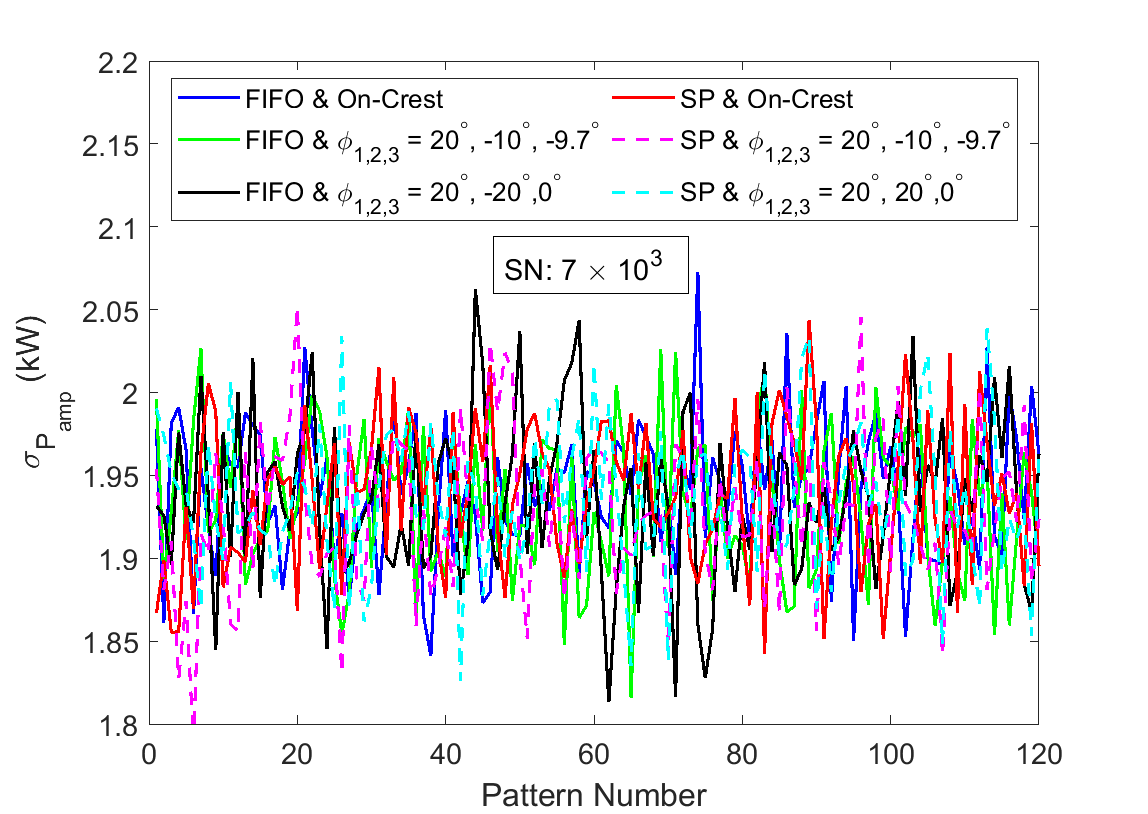}
		\caption{}
	\end{subfigure}
	\caption{Comparison of SP and FIFO at dynamic set-points when beam loading is at on-crest: (a) cavity voltage jitters; (b) cavity phase jitters; (c) average amplifier power; and (d) amplifier power jitters.}
	\label{fig:23}
\end{figure*}

Simulations were performed for SP with on- and off-crest beam loadings and static and dynamic set-points. The results are overlaid for comparison and given in Fig.~\ref{fig:22} and Fig.~\ref{fig:23}. The $S/N$ was set to $7\times10^{3}$ to observe the behavior of the system with moderate noise. Fig.~\ref{fig:22} shows results with on-crest beam loadings only. Fig.~\ref{fig:23} shows results with dynamic set-point only.

\subsection{Comparison of dynamic and static set-points of FIFO and SP}
In the sub-figure (a) of Fig.~\ref{fig:22}, we see SP can have slightly lower cavity voltage jitter $\sigma_{V_{cav}}$ than FIFO. The difference in $\sigma_{V_{cav}}$ between of different patterns of SP are not as significant as FIFO. SP is insensitive to set-points regardless of patterns, wile for FIFO is only insensitive at optimal filling patterns (pattern number 60 and 61). 

In the sub-figure (b) of Fig.~\ref{fig:22}, we see the phase jitters are noise dominated and remained low at around $10^{-3}$ degrees. This shows at $S/N$ of $7\times10^{3}$, the phase jitters is negligible for all injection schemes, set-points, and filling patterns. 

In the sub-figure (c) of Fig.~\ref{fig:22}, we see injection schemes, filling patterns, and set-points all can affect the average beam power. Firstly, we see SP requires minimum power regardless of set-points and filling patterns. Secondly, when FIFO is combined with the dynamic set-point, the average power is minimized as well. Thirdly, When FIFO is with static set-point, the filling pattern becomes the most important factor in determining the average power. When the pattern is optimal, the power 14.9~kW is very close to minimum power of 11.3~kW. If one combines FIFO with static set-point and the worst filling pattern, the average power can be as high as 333~kW, which is 30 times of minimum. All these are important factors to consider and optimize when designing ERLs to minimize power consumption.

In the sub-figure (d) of Fig.~\ref{fig:22}, we see $\sigma_{P_{amp}}$ has similar shape as average $P_{amp}$. It is because $\sigma_{P_{amp}}$ is determinant factor for $P_{amp}$. At dynamic set-point, the $\sigma_{P_{amp}}$ is very small at about 2~kW for all patterns and injection schemes, which is consistent with our earlier results. On the other hand, for the static set-point $\sigma_{P_{amp}}$ can range from 10$-$270~kW, depending on the filling pattern and injection schemes. SP with static set-point is significantly better than FIFO with static set-point, except for the optimal patterns of FIFO.

Over all, dynamic set-points is better than static as it causes less jitters and requires less power. When set-point is static, the optimal patterns can lower jitters and power to near the minimum. SP is more stable than FIFO, even when it is with static set-point.

\subsection{Comparison of on- and off-crest}
In the sub-figure (a) of Fig.~\ref{fig:23}, we see off-crest beam loading lowers cavity jitters slightly, which could be due to the fact that at off-crest phases electron bunches take/deposit less energy from/to the cavity than on-crest. In the sub-figure (b), we phase jitters increased more than 1 order of magnitude for off-crest cases. Therefore, off-crest beam loading causes increase in the phase jitters, but the jitters after the increase is still small at $0.1-0.3$ degrees for our parameter settings. We have intentionally set the bunch charge to a high value of 18.4~nC to accentuate the effect of beam loading. Over all, SP has smaller phase jitters than FIFO. There is no difference in average amplifier power and its power jitters.

\section{Conclusion}
\label{section:conclusion}
We studied recirculating ERL beam loading instabilities of different filling patterns under various noises, phases, and injection schemes by combining analytical model with simulations. Simulation results agreed with analytical predictions with some minor differences at very high or very low noises, possibly due to the non-linearity of the system. These studies give us useful insight to ERL beam loading with different filling patterns, LLRF systems, and injection schemes. 

We found filling patterns, S/N, and LLRF set-points are important for maintaining stable cavity voltage and lowering consumed RF power. We identified optimal filling patterns for 6-turn ERL, but our methodology can be applied for finding optimal patterns of other multi-turn ERLs as well. Optimal filling patterns lower cavity voltage jitters and amplifier power significantly. Our studies show that ERL LLRF requires dynamic set-point voltage. The cavity voltage is more sensitive to the filling patterns than noise. The amplifier power jitters is more sensitive to noise than filling patterns. For our setup parameters, when $S/N$ is increased to $7\times10^3$ or more, the average amplifier power can be reduced to minimum of around 11~kW. Lowering noise is critical for lowering the amplifier power. The effect of charge jitters and off-crest beam loading on the cavity voltage and amplifier power are negligible. The off-crest beam loading increased the cavity phase jitters by one order of magnitude, but jitters are still small at around $0.1-0.3$ degrees.

We have also introduced SP and FIFO injection schemes and found they behave differently, depending on the beam loading type (on- and off-crest), set-points, and filling patterns. Over all, SP is more stable than FIFO and requires less power. 

It will be interesting study to investigate BBU instability for different filling patterns. This work has been done only for 6-turn ERLs, but the theoretical construct and simulation can also be applied to higher or less turn numbers.

\begin{acknowledgments}
The authors would like to thank Dr. Graeme Burt, Dr. Amos Dexter, and Dr. David Walsh for their useful suggestions and insights. The studies presented have been funded by STFC Grants No. ST/P002056/1 under the Cockcroft Institute Core Grant.
\end{acknowledgments}

\bibliography{bibtex}

\end{document}